\documentstyle{mn}
\input {psfig.sty}

\def \bh {_{\rm BH}}
\def \mbh {M_{\rm BH}}
\def \blr {_{\rm BLR}}
\def \ms {$M-\sigma$}
\def \rl {$r-L$}
\def \mgii {Mg\,{\sc ii}}
\def \civ {C\,{\sc iv}}
\def \hbeta {H$\beta$}
\def \disp {{\rm Disp}}

\def \pmo {$^{-1}$}

\def \empha {({\emph a})}
\def \emphb {({\emph b})}
\def \emphc {({\emph c})}
\def \emphd {({\emph d})}
\def \sn {{\rm S/N}}
\def \ed {_{\rm Edd}}
\def \msun {M$_{\odot}$}

\title[Constraining quasars]{Constraining the quasar
  population with the broad-line width distribution}

\author[S. Fine et al.]
       {S. Fine$^{1}$\thanks{sfine@physics.usyd.edu.au},
	 S.~M. Croom$^1$, P.~F. Hopkins$^2$, L. Hernquist$^2$,
	 J. Bland-Hawthorn$^1$,
	 \newauthor M. Colless$^3$, P.~B. Hall$^4$,
	 L. Miller$^5$, A.~D. Myers$^6$, R. Nichol$^7$, K.~A. Pimbblet$^8$,
	 \newauthor N.~P. Ross$^{9,10}$, D.~P. Schneider$^{10}$,
	 T. Shanks$^{9}$ and R.~G. Sharp$^3$ \\
$^1$School of Physics, University of Sydney, NSW 2006, Australia. \\
$^2$Harvard-Smithsonian Center for Astrophysics, 60 Garden Street,
       Cambridge, MA 02138, USA.\\
$^3$Anglo-Australian Observatory, PO Box 296, Epping, NSW 1710,
      Australia.\\
$^4$Department of Physics and Astronomy, York University, 4700 Keele
      Street, Toronto, ON M3J 1P3, Canada. \\
$^5$Department of Physics, Oxford University, Keble Road, Oxford, OX1
      3RH, UK.\\
$^6$Department of Astronomy, University of Illinois, Urbana, IL,
       USA. \\
$^7$Institute of Cosmology and Gravitation, Mercantile House, University
       of Portsmouth, Portsmouth PO1 2EG. \\
$^8$Department of Physics, University of Queensland, Brisbane, Qld
      4072, Australia. \\
$^{9}$Department of Physics, Durham University, South Road, Durham DH1
      3LE, UK.\\
$^{10}$Department of Astronomy and Astrophysics, The Pennsylvania State
      University, 525 Davey Laboratory, University Park, PA 16802, USA.
}

\begin{document}

\maketitle

\begin{abstract}

In this work we test the assertion that the scatter in the mass of
black holes which drive quasars should be luminosity dependent with
less scatter in more luminous objects. To this end we measure the
width of the \mgii\ $\lambda2799$ line in quasar spectra from the
SDSS, 2QZ and
2SLAQ surveys and, by invoking an unnormalised virial mass estimator,
relate the scatter in line width to the scatter in mass in the underlying
black hole population. We find conclusive evidence for a trend such
that there is less scatter in line width, and hence black hole mass, in
more luminous objects.

However, the most luminous objects in our sample show such a low degree of
scatter in line width that, when combined with measures for the
intrinsic scatter in the radius-luminosity relation for the broad-line
region in active galaxies, an inconsistency arises in the virial
technique for estimating black hole masses. This analysis implies that,
at least for the most luminous quasars, either there is little-to-no intrinsic
scatter in the radius-luminosity relation or the \mgii\ broad emission line
region is not totally dominated by virial velocities.

Finally we exploit the measured scatter in line widths
to constrain models for the velocity field of the broad-line
region. We show that the lack of scatter in broad line-widths for
luminous quasars is inconsistent with a pure planar/disk-like geometry for
the broad-line region. In the case of a broad-line region with purely
polar flows the opening angle to luminous quasars must be less than
$\sim55^{\circ}$. We then explore the effects of adding a random or
spherically symmetric component to the velocities of gas clouds in the
broad-line region. Assuming an opening angle to quasars of
$45^{\circ}$ a planar field can be made consistent with
our results if $\sim40-50$\,\% of the velocities are randomly
distributed.

\end{abstract}


\section{Introduction}

It is now apparent that the majority of massive galaxies harbour
a super-massive black hole (SMBH) at their centre. Dynamical studies
of the sphere of influence of these SMBHs have been successful in
determining the mass for some tens of systems. In these cases the mass
of the central SMBH has been observed to correlate strongly with
properties of their host spheroid such as luminosity \cite{mag98},
velocity dispersion ($\sigma$; Ferrarese \& Merritt 2000, Gebhardt et
al. 2000) and concentration/Sersic index \cite{gra01}.  More
recently, Hopkins et al. (2007a) have demonstrated that these
relationships can all be regarded as various projections of 
a ``black hole fundamental plane'' (BHFP), relating black hole mass
to the potential well of the galaxy.

The existence of these correlations and, in particular, the similarity
of the BHFP to the fundamental plane of elliptical galaxies points to
an intimate link between the growth of SMBHs and galaxy evolution.
Moreover, the BHFP is consistent with the notion that SMBHs grew in a
self-regulated manner, from gas gravitationally confined in the galaxy
centre, which was eventually expelled by feedback processes (Hopkins
et al. 2007b). Recent hydrodynamical simulations that incorporate radiative
cooling, star formation, black hole growth and feedback from both
supernovae and nuclear activity have shown that mergers between
gas-rich galaxies of comparable mass provide a mechanism for
concentrating gas in galaxy centres through tidal effects (e.g. Barnes
\& Hernquist 1991, 1996), fueling the growth of SMBHs (e.g. Di Matteo
et al. 2005), reproducing the correlations (e.g. Robertson et
al. 2006), and explaining observed properties of quasars as a phase of
evolution prior to the termination of black hole accretion
(e.g. Hopkins et al. 2005a,b,c).

Further evidence for a merger-driven origin of quasars is provided by
comparing the evolution of the abundance of luminous quasars with that
of the cosmic star formation rate (SFR).  From an empirical
determination of the bolometric quasar luminosity function, Hopkins,
Richards \& Hernquist~(2007) infer that the quasar luminosity density
peaks at $z=2.15 \pm 0.05$ and rapidly declines towards higher
redshifts.  Some observational estimates suggest a similar behaviour
for the evolution of the cosmic SFR (e.g. Fan et al.~2001; Hopkins \&
Beacom~2006).  However, at high redshifts, incompleteness, cosmic
variance of the surveys, and uncertain corrections owing to dust
extinction complicate this analysis.  From a measurement of the
opacity of the Lyman-alpha forest (Faucher-Giguere et al. 2007a),
Faucher-Giguere et al. (2007b) have shown that the optical depth at $z
= 4$ is incompatible with a steep decline in the cosmic SFR at $z >
3$, as suggested by e.g. the results of Hopkins \& Beacom~(2006), but
in accord with theoretical modeling (e.g. Springel \& Hernquist 2003;
Hernquist \& Springel 2003) which predicts that the SFR should peak at
$z>4$.  The implied offset between the peak in the cosmic SFR and the
quasar luminosity density indicates that the growth of SMBHs is not
directly tied to star formation or gas density, but is related to a
secondary process.  By employing estimates of the rate of halo
mergers, Hopkins et al. (2007c) have shown that the quasar luminosity
function can be reproduced if SMBH growth occurs primarily in major
mergers of gas-rich galaxies.

Motivated by these various lines of argument, Hopkins et al. (2006a,b,
2007a,b) have developed a model for galaxy evolution in which mergers,
starbursts, quasars, SMBH growth, and the formation of ellipticals are
connected through an evolutionary sequence.  As part of this work,
Hopkins et al. (2005a) used simulations of merging galaxies to
quantify the phases of evolution associated with quasar activity and
showed, in particular, that quasar lifetimes depend not only on the
instantaneous luminosity of a quasar, but also its `peak'
luminosity. Convolving these lifetimes with estimates of the merger
rates of galaxies, Hopkins et al. (2005b,c; 2006a; 2007a) were able to
reproduce the observed optical and X-ray luminosity functions of
quasars.

In this interpretation of the quasar luminosity function, the brighter
($>L^*$) objects are all massive black holes, accreting near the
Eddington rate towards the end of their growth phase. Less luminous
quasars can be either low-mass systems accreting rapidly and
undergoing significant growth, or larger black holes accreting at
comparatively lower rates.

This model implies that the range of accretion rates in quasars
should be luminosity dependent. That is, the range accretion rates
(and by extension SMBH masses) at a given luminosity should be larger
for lower luminosity objects, and decrease as the luminosity increases.

In this work we aim to derive the dispersion of the SMBH mass of
quasars as a function of luminosity. We measure the \mgii\
broad emission-line width in spectra from three large spectroscopic
surveys of Type I active galactic nuclei (AGN). Assuming there exists 
a virial relation for calculating SMBH masses from the \mgii\ line
width we relate the measured
distribution of emission line widths to the SMBH mass distribution.

\section{Virial SMBH mass estimators}

By far the most accurate, robust and believable method for measuring
the mass of SMBHs is to perform dynamical studies of stars, gas
or masers in the potential of the SMBH (e.g. Herrnstein et al. 1999;
de Francesco et al. 2006; Gebhardt et al. 2003). Rarely, however, is this
viable for type I AGN which for the most part are too
distant to resolve the sphere of influence of the SMBH, and have nuclei
so bright that it is difficult if not
impossible to obtain detailed photometric or spectroscopic information
of the environment at redshifts $\sim1$.

The most direct method for calculating SMBH masses of type I AGN is
via reverberation mapping of the broad-line region (BLR)
and the virial theorem \cite{pea04}. In this case the size of the BLR
($r\blr$) is given by the time delay between continuum and emission
line variations and the velocity dispersion of the BLR ($V\blr$) is
measured as the width of the emission line in the variable
spectrum. The mass of the SMBH is then estimated as 
\begin{equation}
M\bh=f\frac{r\blr V^2\blr}{G}.
\label{equ_virial_real}
\end{equation}

The factor $f$ defines what is not known about the BLR; its
geometry, velocity field and orientation. The value of $f$ is order
unity and authors take various approaches assigning it a value. One
can find $f=3/4$ applicable to random orbits \cite{p+w99},
$f=1/(4\sin^2\theta)$ for a disk inclined at an
angle $\theta$ to the observer \cite{m+d01}, or even $f=1$ for simplicity
\cite{m+j02}. Onken et al. (2004) measured $\overline{f}=1.4$ by
comparing the \ms\ relation in a group of reverberation mapped AGN
in local spheroids.

This last value is somewhat higher than
expected compared to the simple theoretical values. However,
the interpretation of this is unclear and the discrepancy may be due
to selection bias \cite{lau07} and/or cosmological evolution in the
\ms\ relation \cite{woo06} both of which could bias the measured value
of $f$ high.

Due primarily to time constraints, the number of reverberation mapped
systems is in the tens and the luminosity range which these
measurements span is not huge \cite{kasp07}. However, one key result
which has come out of reverberation mapping is the radius-luminosity
relation \cite{kasp05}. While
there is significant scatter in this relation, including significant
intrinsic scatter of up to 40\,\%, it does offer a simple
single epoch method for estimating the radius of the BLR in AGN.

If one takes the instantaneous FWHM of a broad-emission line as a
measure of the virial velocity in the BLR, then the \rl\ relation
provides a method for estimating the unscaled SMBH mass with single epoch
observations. This technique has become known as the virial method.
To date virial relations have been calibrated for the \hbeta\
\cite{kasp00,v+p06}, \mgii\ \cite{m+j02,m+d04,kol06,sal07,mcg08} and
\civ\ lines \cite{vest02,v+p06}. In each case
these calibrations result in expressions of the form
\begin{equation} \label{equ_virial}
M\bh=A(\lambda L_{\lambda})^{\alpha}FWHM^{2}.
\end{equation}
Where $FWHM$ is the full width at half maximum of the spectral line in
question and $\lambda L_{\lambda}$ is the monochromatic luminosity of
the continuum near that line. $A$ is a normalisation constant and the
exponent $\alpha$ gives the luminosity dependence of the \rl\
relation.

\subsection{Problems with virial  mass estimators}

When using virial SMBH masses it is prudent to state some
caveats. Firstly and most obviously the virial relations are
statistical in nature. Hence while they may be accurate when averaged
over a large number of systems individual measurements should be
viewed with caution.

Secondly a number of studies have investigated the differing
emission regions for high and low ionisation lines
(e.g. Richards et al.~2002; Baskin \& Laor~2004; Elvis~2004 which
raises questions as to whether all
broad lines can be used as virial mass indicators. This is,
however, still a subject open to debate. Vestergaard \& Peterson (2006)
reanalysed Baskin \& Laor's data
and showed that the apparent discrepancies between \hbeta\ and
\civ\ virial mass estimates are nullified by applying more appropriate
selection criteria.

Thirdly it is important to note that the calibrations of the \mgii\
and \civ\ virial relations are not direct. These lines have not been
sufficiently studied in reverberation mapping campaigns to calibrate
any potential \rl\ relation for them. Instead the virial relations for
these lines are normalised through comparisons with SMBH masses
calculated from the \hbeta\ line.

Finally, there is simply a dearth of solid information
constraining the velocity field and geometry of the BLR. Reverberation
mapping as a technique has the potential to describe the BLR in detail
given sufficient quality data \cite{w+h91}. However, in reality this
sort of idealised precision is unlikely and to date these types
of results have not materialised. Theoretical models for BLRs exist
(e.g. Emmering, Blandford \& Shlosman 1992; K\"{o}nigl \& Kartje 1994;
Murray \& Chiang 1996) but the lack of strong observational
constraints makes a proper comparison between these difficult.
However, almost all of these models imply a strong virial component to the
velocity field of the BLR.

There is good evidence that the virial method, while
imprecise, can on average give accurate black hole masses. Comparisons
between reverberation masses and bulge velocity dispersion show an
\ms\ relation analogous to that observed in nearby quiescent galaxies
\cite{onk04}. And comparisons between virial and reverberation masses
also show agreement \cite{v+p06,m+j02}. But to obtain strong results
from virial masses large samples are required to beat down random errors in
these estimates.

In this work we assume a virial relation for the
\mgii\ line as given by equation~\ref{equ_virial}. We do not, however,
assume a parametrisation for this relation since it is not required.
We take a large number of QSO spectra
from several spectroscopic surveys and bin them by luminosity and redshift.
We invoke the \rl\ relation and assume objects in a luminosity bin
will have the same $r\blr$, and the 
scatter in SMBH masses within that bin is simply given by the scatter in
broad-line widths. Furthermore, looking at the scatter in log space
for sufficiently small luminosity bins
we can ignore the $L$ term in equation~\ref{equ_virial}, and the
coefficient and
\begin{equation} \label{equ_llw_disp}
\disp(\log(M\bh))=2\times\disp(\log(FWHM_{\rm \mgii}))
\end{equation}
Where Disp() denotes the dispersion in the given variable.
Note that we have ignored all extraneous sources of scatter in this
equation, some of which will be significant to our calculations. We
discuss these in more detail in sections~\ref{sec_virial_problems}
and~\ref{sec_blr_geom}.


\section{Data}

We take as our sample all of the quasar spectra from the Sloan Digital
Sky Survey (SDSS; York et al.~2000) data release five (DR5;
Adelman-McCarthy et al.~2007)
as compiled by Schneider et al. (2007). To increase our
luminosity range we also take all QSO spectra from the 2df QSO
Redshift survey (2QZ; Croom et al. 2004)
and 2dF SDSS LRG And QSO survey (2SLAQ; Richards et al. 2005; Croom et
al. in prep.) as well. Table~\ref{tab_surveys} shows a
brief summary of the number of objects and magnitude limits in each sample.

\begin{table*}
\begin{center}
\caption{Summary of the surveys from which we obtained
  spectra. Successive surveys have fewer spectra but go deeper,
  increasing our luminosity range at a given redshift. Note that the
  magnitude limits quoted for the SDSS QSO survey are those for the
  primary QSO survey. The high redshift sample goes deeper, and
  included are sources observed under different selection criteria and
  also QSOs identified as part of other surveys.}
\label{tab_surveys}
\begin{tabular}{cccccc}
\hline \hline
Survey      & No. of Objects & Mag. Limits & Resolution & Dispersion &
  $\overline{\rm S/N}$ \\
\hline
SDSS (DR5)  & 77,429 & $19>i>15$               & $\sim165$\,km/s &
  $\sim1.5$\,\AA\,pix\pmo & $\sim13$\,pix\pmo  \\
2QZ         & 23,338 & $20.85>b_{\rm J}>18.25$ & $\sim465$\,km/s &
  $\sim4.3$\,\AA\,pix\pmo & $\sim5.5$\,pix\pmo \\
2SLAQ       &  8,492 & $21.85>g>18.00$         & $\sim465$\,km/s &
  $\sim4.3$\,\AA\,pix\pmo & $\sim5.5$\,pix\pmo \\
\hline \hline
\end{tabular}
\end{center}
\end{table*}

We thus have spectra from three different surveys taken with two
different instruments. We give here a brief description of the spectra.

\subsection{SDSS spectra}

Details of the Sloan telescope and spectrograph are given in Gunn et
al.~(2006) and Stoughton et al.~(2002).
The spectra have a logarithmic wavelength scale
translating to a dispersion of
$\sim1-2$\,\AA\,pix\pmo\ and a
resolution $\lambda/\Delta\lambda\sim1800$ in the wavelength
range $3800-9200$\,\AA. Objects are observed initially for
2700\,sec. Then are reobserved in 900\,sec blocks until the median
S/N is greater than $\sim4$\,pix\pmo resulting in a S/N distribution
with a mean at $\sim13$\,pix\pmo.

The spectra are extracted and reduced with the {\sc spectro2d}
pipeline and automatically classified with {\sc spectro1d}
\cite{sto02}. However,
in creating the Sloan QSO sample used in this paper, Schneider et
al. (2007) visually inspect all of the candidate spectra to
determine their classification.


\subsection{2dF spectra}

Both 2QZ and 2SLAQ spectra were taken with the 2 degree Field (2dF)
instrument on the Anglo-Australian Telescope with the 300B grating
\cite{lewis02}. Spectra have a
dispersion of 4.3\,\AA\,pix\pmo\ and a resolution of $\sim9$\,\AA\ in
the wavelength range $3700-7900$\,\AA. 2QZ
observations were between 3300 and 3600\,sec compared with 14400\,sec
for 2SLAQ. The increase in exposure time for the fainter 2SLAQ
sample results in S/N distributions that are almost
indistinguishable. Both peak at $\sim5.5$\,pix\pmo\ for positive QSO
IDs. 2dF spectra are extracted and manually classified during the
observing run with the 2dFDR pipeline \cite{b+g99} and {\sc autoz}
redshifting code \cite{croom01}.

\begin{figure}
\centering
\centerline{\psfig{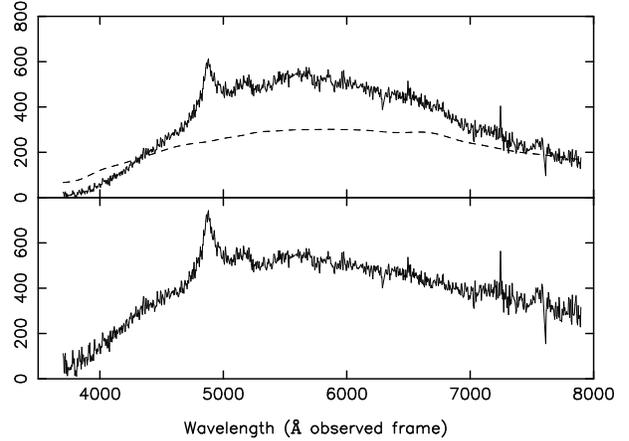}}
\caption{Example of a 2dF spectrum with poor response at the blue
  end. In the top panel the original spectrum is shown as reduced by
  2dFDR, the dashed line shows the response correction from Lewis et
  al. (2002) scaled for comparison. In the bottom panel this
  correction is applied to the spectrum. We find
  that, while this does improve the shape of the
  spectrum, it is inadequate in this case to correct for the sharp
  drop in received flux at the blue end.}
\label{fig_badspec}
\end{figure}

The main difference between reduced Sloan and 2dF spectra
is the lack of flux calibration for 2dF sources. An average flux
calibration for the 300B grating has been calculated by Lewis et
al. (2002) as part of the 2dF Galaxy Redshift Survey and
in our analysis we do apply this correction. However, a quick
examination of 2dF spectra shows the inadequacy of this median
calibration to correct for the variations in response between
differing spectra as illustrated in Fig.\ref{fig_badspec}.

\section{Analysis}

Our goal is to measure the dispersion in the widths of broad \mgii\ lines
in a large sample of QSOs. For this we require an automated routine to
measure the width of \mgii\ in a consistent way across our sample and
a robust method for analysing the results. In this section we discuss
the line fitting code used to measure the line widths, and our
analysis of the output, but first a quick note on absolute magnitudes.

\subsection{Magnitude calculations}

Magnitudes are given in the $b_J$-band in the 2QZ catalogue, and Sloan
$ugriz$-bands in both the SDSS and 2SLAQ catalogues. Richards et
al.~(2005) compared the $b_J$ and $g$ pass bands and found them to be
roughly equivalent. Examining 2QZ QSOs with $g$-band imaging they
found a consistent $\langle g-b_J\rangle=-0.045$. In our analysis we
use $g$-band 
magnitudes for all objects taken from the SDSS and 2SLAQ datasets, and
for the 2QZ we apply this correction to their $b_J$ magnitudes.

Throughout
this paper we use a flat $(\Omega_{\rm m},\Omega_{\Lambda})=(0.3,0.7)$,
$H_{0}=70\,{\rm km\,s}^{-1}\,{\rm Mpc}^{-1}$ cosmology when
calculating absolute magnitudes. We use the K-corrections laid out
by Cristiani~\&~Vio~(1990) and correct for galactic extinction as
advised in the relevant catalogue paper.

\subsection{Line measurements}

Our spectral analysis routine implements two separate fitting processes
which are linked via iteration. Firstly a combined iron and
continuum model is fit to the region of the spectrum not affected by the
\mgii\ line, then a Gaussian profile is fit to the line itself. We
iterate the procedure to more accurately define the region of the
spectrum affected by \mgii\ emission, and so improve our iron and
continuum fits. This significantly improves our spectral line
calculations.

\subsubsection{Continuum and iron correction}

\begin{figure*}
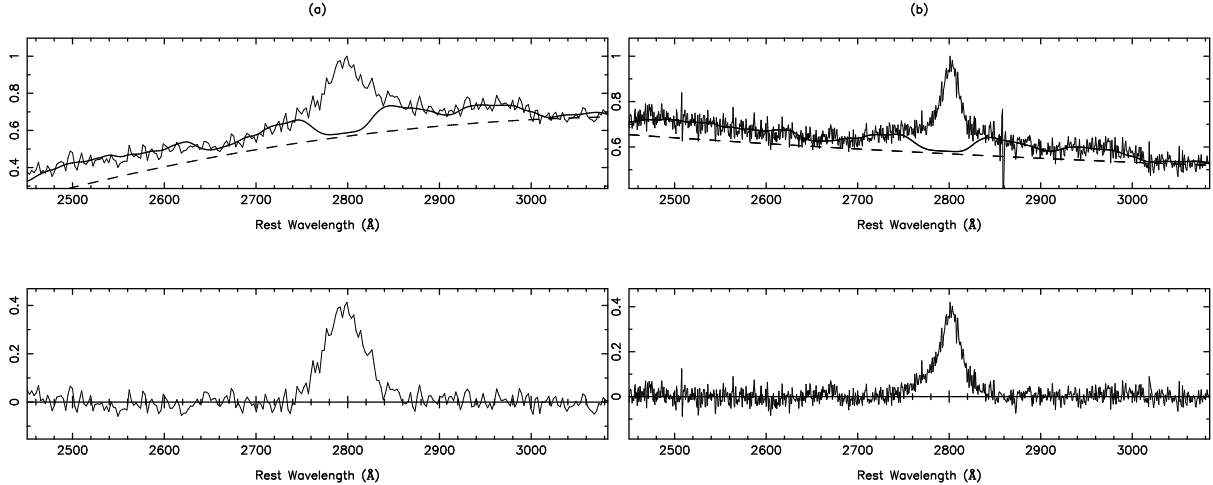

\centering
\centerline{\psfig{file=cont_fit_2qz.ps,width=8.0cm,angle=-90}\psfig{file=cont_fit_sdss.ps,width=8.0cm,angle=-90}}
\caption{These figures illustrate the iron and continuum fitting
  process on two very different spectra. In each case the top panels
  show the \mgii\ line region in the original spectrum, heavy dashed
  and solid lines show the continuum and iron + continuum model fit to
  it respectively. The bottom panel then shows the residual when this
  is subtracted from the spectrum. \empha\ shows the fit to
  the spectrum of J095421.6-000152 as observed as part of the
  2QZ. This is the same spectrum as given in Fig.\ref{fig_badspec}
  and shows a strong turnover in received flux at the blue
  end. \emphb\ shows the spectrum of J000042.89+005539.5 as observed
  with the Sloan telescope and exhibits the classic quasar power law
  continuum. For both cases the quadratic continuum we use produces an
  excellent fit to the data. In all plots the y axis gives the
  normalised flux.}
\label{fig_cont_fits}
\end{figure*}

Continuum emission from QSOs is
well described by a power law in the optical-UV region. However, 2dF
spectra are not flux calibrated. Variations in response modulate
the shape of 2dF spectra most often characterised by a
drop off in flux at the blue end. Since we cannot accurately model these
spectra with a power law we describe the continuum with a quadratic
fit. This can then simultaneously correct for response effects in 2dF
spectra and, since we only fit this to a limited region of the
spectrum, approximate a power-law shape for Sloan
objects. Fig.\ref{fig_cont_fits} shows our iron and continuum fits to
both a 2dF spectrum which has a drop off in flux at the blue end, and
a well-calibrated SDSS spectrum.

To remove iron emission from our spectra we use the template for QSO
iron emission derived from the narrow-line Seyfert I object I Zw 1 by
Vestergaard \& Wilkes (2001). This template is derived from an
intrinsically narrow-line object which made it possible to isolate the
species responsible for each component of the emission. Unfortunately,
because the template is derived from a real object, it is least well
defined in the region around strong emission lines where de-blending
differing species becomes difficult.

This limitation is of particular importance directly beneath the \mgii\ line
where the template shows no iron emission. This is not due to a quantified
lack of emission, but to the difficulty of de-blending the iron and 
\mgii\ emission in this region. It has now become common practise to add
flux to the template in this region (e.g. Kurk et al. 2007; Salviander
et al. 2007) to make it more
consistent with theoretical models \cite{s+p03}. We follow the method of
Kurk et al.~(2007) and add a constant level of flux under the \mgii\
line at 20\,\% the intensity of the average in the
$2930-2970$\,\AA\ (rest frame) region. We find that the level of flux included
under the \mgii\ line does affect our results. Given the
range of likely levels of iron emission in this region, however, the affect is
not significant and does not impact our conclusions. This will be
discussed further in section~\ref{sec_res_Fe}.

Since the iron template is derived from a narrow-line object it
must be smoothed to properly describe the iron emission in a
broader-line QSO. We follow the same procedure as previous authors
(e.g. Boroson \& Green 1992) and make a selection of iron templates
smoothed by Gaussians of width 500, 750, 1000, 1500, 2500, 5000,
7500 and 10,000\,km/s. We fit all of these templates to this spectrum
and accept the best fit in terms of the $\chi^2$.

We fit for continuum and iron emission simultaneously with the {\sc
svdfit} routine \cite{press89}. The fit is performed on a region
bounded at the blue end at 2450\,\AA\ by the [Ne\,{\sc
iv}] $\lambda2424$ line, and at the red end at 3085\,\AA\ which is the
limit of the iron template (all wavelengths are rest frame). We also
mask the \mgii\ line out of our fitting.

On the first attempt we use a fiducial $\pm50$\,\AA\ mask for
the line but on subsequent iterations we use a Gaussian fit to the
line to define its boundaries.

It is important to keep in mind that once we have subtracted
the iron and continuum contributions from the data we introduce a strong
covariance into our spectrum and individual pixel values can no longer
be thought of as independent in subsequent error calculations.

\subsubsection{Gaussian fitting, iteration and line width calculations}
\label{sec_gfiting}

We subtract the iron and continuum contribution to the spectrum and
fit a single Gaussian to the \mgii\ line with the {\sc mrqmin}
routine \cite{press89}. We then perform the iron and continuum fit to the
original spectrum again masking out data within $\pm1.5FWHM$ of the centre
of the fitted Gaussian.
This process is repeated until successive Gaussian fits have FWHMs
consistent to within $1/2$ their error.

In this analysis some thought must be given to what
statistic to use when defining the width of a spectral
line. The width of single and/or multiple Gaussian/Lorentzian fits to
the line are often used to describe the line width (e.g. McLure \&
Dunlop 2004; Shen et al. 2007; Wang,
Lu \& Zhou 1998). We have already performed Gaussian
fits to our line and we do find that despite the clear non-Gaussian
nature of the \mgii\ line these give a reliable measure for the
width when compared with the other methods outlined here. However, we
find non-parametric estimators for the line width
more appealing for this analysis, primarily because they make no
assumptions as to the line profile, but also because their errors can
readily be calculated including the contribution of covariance
introduced into the spectrum in our iron and continuum correction.

A common statistic used for describing line width is the
FWHM \cite{vest02,v+p06}. In high S/N spectra this gives a
good determination of the line width, and the error for the FWHM can
readily be calculated including the contribution from covariance in
the spectrum. Unfortunately the measured FWHM is quite susceptible to
noise. How to define the maximum flux density of a line in an unbiased way
is unclear and multiple crossings of the half maximum value in noisy
spectra demand some sort of averaging which can also bias the
measurement. Further we find that the susceptibility of the FWHM to the
values of a relatively small number of pixels in a spectrum makes
it an unstable and often inaccurate measure of the line's true width.

Another measure of the line width that is becoming increasingly
widespread is the dispersion of the line $\sigma$
\cite{f+m00,v+p06,wil07}. However, we find that the excessive weighting this
measure gives to the values of pixels in the wings of the lines makes
it an unreliable estimator of line width in low S/N spectra.

Finally inter-percentile values (IPV) can be measured
\cite{whit85}. While at first
glance the process of measuring an IPV width is similar to measuring
the FWHM,
the dependence of IPV widths on the cumulative flux distribution rather
than the flux density at a given point makes the IPV measurements
considerably more
robust. Like the dispersion, IPV widths are somewhat affected by noise in the
wings of lines; in particular this can affect the total flux of a
line and how one defines the zero-point of the cumulative flux
distribution. However, when calculating the dispersion the weight
given to a single pixel is proportional to the square of the
displacement of that pixel from the line centre. This power-of-two
dependence makes $\sigma$ very susceptible to noise in the wings of a
line which is not a problem for IPV widths.

Our calculation for the IPV width is performed only on the part of the
spectrum within $\pm1.5FWHM$ of the Gaussian fit to the \mgii\
line. In this region we find the cumulative flux distribution and
search this for the first crossing points at $1/4$ and $3/4$ the total
flux. We then interpolate between the adjacent pixels
either side of the crossing to obtain sub-pixel accuracy in our IPV
estimate.

Errors on the IPV widths are calculated from the spectral variance
array including the contribution of covariance introduced by the iron
and continuum subtraction. We find the mean error of these measurements
to be $\pm0.05$\,dex.

Finally we subtract the resolution of the spectrograph in quadrature
from our measured line width under the assumption of a Gaussian
profile for both the emission line and instrumental resolution. Since
the \mgii\ line is not Gaussian in shape this is only an approximate
correction and to test its validity we evaluate the effect of smoothing
Lorentzian profiles by a Gaussian. In the worst case scenario of a
Lorentzian line with the smallest IPV width we measure in our sample
($\sim1200$\,km/s; see Fig.~\ref{fig_lw_dist}), smoothed by a Gaussian
with FWHM 675\,km/s (equivalent to the 9\,\AA\ resolution of the 2dF
spectrograph at $\lambda=4000$\,\AA) the correction above
results in an overestimate of the line width by $<5$\,\%. For broader
spectral lines, and in cases where the spectral resolution is greater,
this effect is diminished and hence we make no further attempt to
correct for instrumental smoothing.

\subsubsection{IPV widths vs. FWHM}

In taking the IPV width as our measure of line width we assume there
is a linear scaling relation between the differing
line width parameterisations. Specifically we assume that
equation~\ref{equ_llw_disp} holds equally true for IPV
widths. To test this assumption we also measure the FWHM of the \mgii\
lines in our sample to compare with the IPV widths.

\begin{figure}
\centering
\centerline{\psfig{file=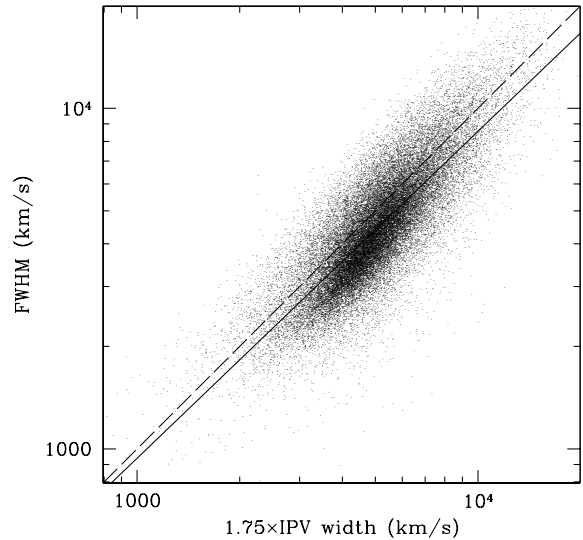,width=8.0cm}}
\caption{Comparison of 50\,\% IPV widths anfd FWHMs for our
  sample. The dashed line gives the 1:1 relation and the solid line
  shows the y on x least squares fit. IPV widths have been multiplied
  by 1.75 for the comparison which is the expected ratio for a
  Gaussian line}
\label{fig_ipv_fwhm}
\end{figure}

These are measured by defining the maximum flux density of the
spectral line as the average of the highest pixel value and the two pixels
adjacent to it. Then the lower and upper bounds of the FWHM are found
by averaging all crossings of the half maximum value on the red and
blue sides of the lines center. Fig.~\ref{fig_ipv_fwhm} shows a
comparison between the IPV width and FWHM measurements for our
dataset. In the figure we have multiplied the IPV widths by 1.75, the
ratio between IPV and FWHM for a Gaussian. There is considerably more
scatter in the FWHM direction due to the susceptibility of this
statistic to noise in the spectrum. However, a y on x least squares
fit to this data gives
\begin{equation}
\log(FWHM)=0.98\log(IPV) + 0.1,
\end{equation}
a linear relation between the IPV width and FWHM. The offset is due
due to the profile of the \mgii\ line which tends to have more flux
in the wings of the line compared to a pure Gaussian.

\subsubsection{Rejection criteria}

With an automated routine
it is impossible to accurately fit every spectrum and there is always
the chance of catastrophic failure. We make a series of checks
and cuts to the results to try and reduce the number of
spurious fits in our final sample.

Failure to produce a satisfactory fit is normally due to a
property of the spectrum being analysed. Absorption features, low S/N,
extreme curvature in the underlying spectrum, bad pixels and residual
telluric features can all cause failures, and commonly a failure is due to a
combination of these.

To test the effect of S/N on the accuracy of our fitting routine we
simulate large numbers of noisy spectra by adding Gaussian noise to
high quality Sloan and 2dF spectra. We then measured the width of the
\mgii\ line in these degraded spectra and compared our results with
the measurement from the high S/N spectrum.

Down to a $\sn\sim1.5$\,pix\pmo\ we find no systematic deviation between the
average line width measured from noisy spectra and the true line
width. Furthermore we find that above this S/N the errors on our measurements
correctly describe the scatter we observe in the measured line widths. We
therefore apply a spectral S/N cut of $\sn>1.5$\,pix\pmo\ to our sample.

It is worth noting that while our error calculations assume Gaussianity
in linear space we find that the distribution of IPV widths measured for
a given spectrum at low S/N is better described by a log-normal. This
may not be
surprising since the widths are randomly distributed and
limited to be $>0$, however, that our errors transfer from linear to
log space is encouraging. We transform the linear error $\sigma_{lin}$ to
the log error $\sigma_{log}$ via its ratio to the measured line width
$IPV$. If we propagate the error through the log then
\begin{equation}
\sigma^2_{log} = \left(\frac{\partial}{\partial
  IPV}(\log_{10}(IPV))\right)^2\sigma^2_{lin}
\end{equation}
\begin{equation}
\sigma_{log} = \frac{\sigma_{lin}}{IPV\ln(10)}.
\label{equ_error_prop}
\end{equation}

We make three further constraints on our data as to whether we will
accept a particular fit. We apply a redshift cut to avoid
contamination by the many telluric features towards the red end of our
spectra, we limit the number of times we iterate our
fitting procedure as described in section~\ref{sec_gfiting}, and
finally we try to eliminate broad-absorption line (BAL) objects which
could contaminate our data.

Telluric features pervade unreduced spectra redwards of
7100\,\AA. While both {\sc spectro2d} and 2dFDR try to remove sky
features the reduced spectra often contain residuals in this region. To
avoid the worst of this we make a simple redshift cut at
$z=1.5$ which ensures that the \mgii\ line and much of our continuum fit
avoid this region.

For a typical spectrum we iterate the fitting procedure outlined
above 2-3 times to obtain convergence. We limit the maximum number of
iterations allowed to 20 and reject any fits which have not converged
by this time as unreliable.

BAL objects are a contaminant in our data and we try to identify and
reject these during our fitting. In the 2QZ and 2SLAQ catalogues many
of the most severe BAL objects are flagged as such and are
not included in our analysis. Those which are not flagged in a
catalogue are identified on the basis that they will have pixels which
are significantly deviant from the rest of the spectrum. After every
Gaussian fit we reject any pixels which deviate by $>3\sigma$
from the fit and we do not include these pixels in subsequent
iterations. Once convergence is obtained in our fitting process we
reject any object which has consecutive rejected pixels spanning
$500$\,km\,s\pmo.

Note that this process of rejecting pixels does not only affect
BAL objects; any pixels with outlying values will be
excised. If this process removes more than one fifth of the pixels
within $\pm1.5FWHM$ of the \mgii\ line we also reject the fit from our
data.

Details of how many objects are rejected due to these various cuts are
given in table~\ref{tab_data}. In total $\sim4$\,\% of the objects are
rejected from our sample. More than 80\,\% of these are objects
rejected due to the fitting routine not converging on a
solution which occurs primarily on low S/N spectra. This, combined
with the imposed S/N cut could create a bias at low luminosity in
our data, in particular if there is a strong correlation between IPV
width and luminosity. However, as we shall see there is only slight
evidence for a correlation between luminosity ind IPV width. In
addition to this the small number of objects rejected mean that any
bias will be negligible. On the other hand the BAL rejection is
designed to expel true outliers from our data and should not bias the
results in any way.


\begin{table*}
\begin{center}
\caption{Table outlining our final dataset. For each of the samples
  we give the total number of objects for which we have attempted to
  analyse the \mgii\ line as well as the apparent
  (uncorrected for Galactic extinction) and absolute
  (extinction corrected) magnitude range of these objects. We
  also give the number of these fits which have been rejected from our
  analysis by each of our criteria. These are low S/N spectra,
  possible BAL objects, and objects
  for which the fitting code did not converge on a result within twenty
  iterations. Note that the redshift cut at $z=1.5$ to avoid sky
  contamination has already been
  factored into the number of spectra in column 2. We also
  give the total number of fits rejected, since an object can be
  rejected for a number of reasons this does not equal the sum of the
  previous 3 columns.}
\label{tab_data}
\begin{tabular}{cccccccc}
\hline \hline
 & & & & \multicolumn{4}{c}{No. of fits rejected} \\
Sample & No. fit & App. Mag. Range & Abs. Mag. range &
 S/N & BAL & Itt.$>$20 & All\\
\hline
2SLAQ & 2684  & $25.09>g>18.12$ & $-19.48>M_{g}>-26.38$ & 44
 & 53   & 142  & 193   \\
2QZ   & 7209  & $20.80>g>18.20$ & $-21.07>M_{g}>-26.74$ & 161
 & 75   & 356  & 425   \\
SDSS  & 23725 & $23.33>g>15.32$ & $-20.98>M_{g}>-29.27$ & 83
 & 141  & 653  & 786   \\
All   & 33618 & $25.09>g>15.32$ & $-19.48>M_{g}>-29.27$ & 288
 & 269  & 1151 & 1404  \\
\hline \hline
\end{tabular}
\end{center}
\end{table*}

\begin{figure}
\centering
\centerline{\psfig{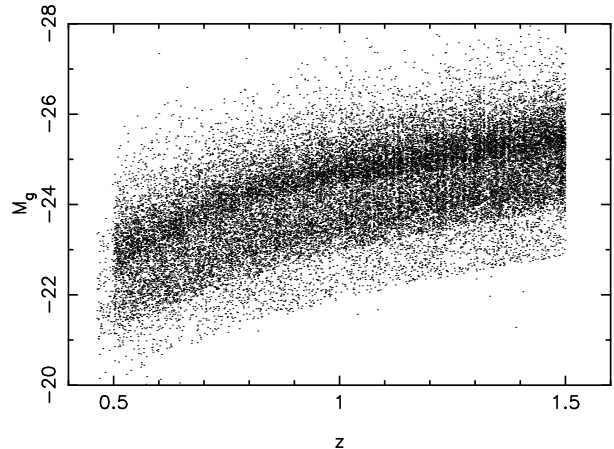}}
\caption{Distribution of our final sample of objects in the
  redshift-luminosity plane. Magnitude limits for the various surveys
  can be made out as lines where the density of objects increases. The
  very few objects with $g$-band magnitudes below the 21.85 cut for
  2SLAQ were selected as potential high redshift objects in the $i$-band.}
\label{fig_M_z_dist}
\end{figure}

\begin{figure*}
\centering
\centerline{\psfig{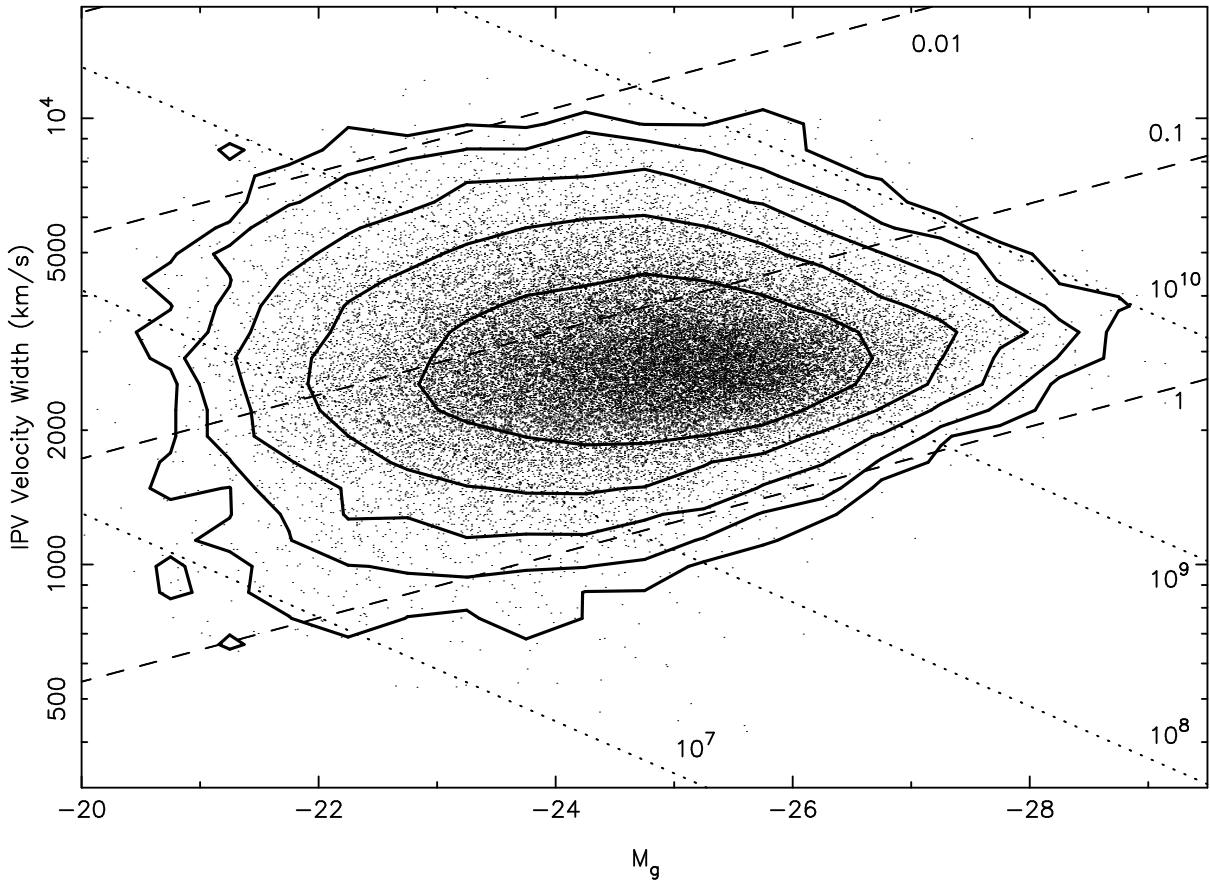}}
\caption{This plot gives the results of our line fitting analysis. We
  plot the absolute $g$-band magnitude of the source
  vs. the measured IPV velocity width of the \mgii\
  line. Over plotted are lines of constant SMBH mass (dotted) and
  Eddington ratio ($L/L\ed$; dashed) as a guide to where these objects
  fall in mass-accretion space. Masses are labeled in units of
  \msun. We have added contours of even density to the plot to
  highlight the shape of the distribution. The contours are evenly
  spaced in terms of log(density).}
\label{fig_M_lw}
\end{figure*}

\vspace{0.5cm}
Fig.\ref{fig_M_lw} shows the results from our fitting plotted on a
line width - magnitude diagram, and Fig.~\ref{fig_M_z_dist} shows how
these objects are distributed on the redshift-luminosity plane. Each
point represents an object with an accepted line width
measurement. The fitting results for these objects is available from
the 2SLAQ website ({\sc www.2slaq.info}).

Added to the plot are lines of constant SMBH mass (dotted) and
Eddington ratio (dashed). These are calculated using the McLure \&
Dunlop~(2004) calibration for the \mgii\ virial relation, assuming
their $B$-band bolometric correction and taking $g\sim B-0.11$ to
calculate the bolometric luminosity of the sources. We then use
a bolometric correction to the continuum luminosity at
3000\,\AA\ of 5.2 \cite{rich06} to calculate the continuum luminosity
at this wavelength from the bolometric luminosity. The McLure \&
Dunlop calibration was based on line widths measured in a different
fashion to this work. They fit two Gaussians to the \mgii\ line and
take the FWHM of the broader component as their line width. We
correct the lines in Fig.\ref{fig_M_lw} for our use of IPV widths
assuming a Gaussian profile for the \mgii\ lines, but McLure \&
Dunlop's use of two Gaussians in their fitting will likely bias the
plotted lines high by a factor of 1.5 to 2. These lines are
plotted more as a guide as to how lines of constant mass and Eddington
ratio would lie on the diagram rather than being exact in normalisation.

We will discuss the implications of this diagram further in
section~\ref{sec_disc}, but already if one compares contours on the left and
right hand side of this 
plot it appears to suggest that there is less scatter in
broad-line widths for more luminous quasars. Furthermore the average line
width shows very little variation across the luminosity range sampled,
a result consistent with previous analyses
\cite{cor03,shen07}.

\subsection{Dispersion analysis} \label{sec_disp_anal}

To investigate how the dispersion in $\mbh$ depends on the
luminosity of QSOs in our sample we bin our sample by luminosity
and calculate the dispersion in line widths in each bin. To remove
possible redshift evolution in this relation we also bin by
redshift. As with measuring line widths some consideration must be
given to choosing the most robust estimate for the dispersion in each bin. 

\begin{figure}
\centering
\centerline{\psfig{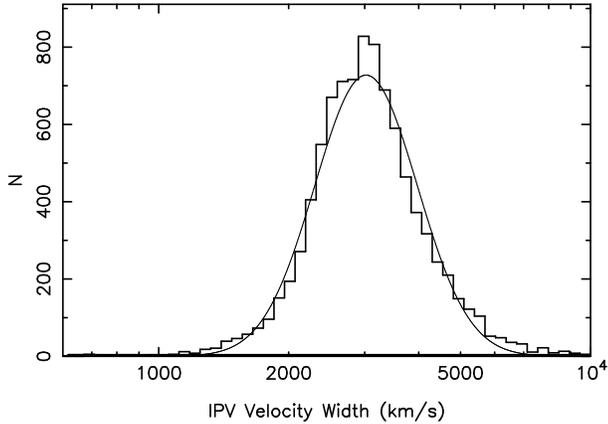}}
\caption{Distribution of \mgii\ line widths for our sample (heavy
  histogram) with a Gaussian fit for comparison (fine line). We find
  this to be roughly log-normal although with slightly more objects in
  the wings in particular towards larger line widths.}
\label{fig_lw_dist}
\end{figure}

The overall distribution of QSO line widths is shown in
Fig.\ref{fig_lw_dist}. We find it is roughly log-normal although with
extended wings and a slight asymmetry. This further motivates our use of
logarithms in equation~\ref{equ_llw_disp} to derive the dispersion in
$\mbh$.

The obvious way to measure dispersion is with the rms in log
space. This approach is, however, very susceptible to outliers.
Objects with either very large or very small results for their line
width are more likely to be due to poor spectral fitting and tend
to have large errors on these measurements. We find that the IPV
method is less biased by these outliers although for the most part we
find very little difference between the two statistics.

We must be wary that the distribution of measured line widths
is not a direct representation of the intrinsic distribution of
line widths in our sample. Instead the measured distribution
represents the intrinsic distribution convolved with the distribution
of errors on these estimates.
In the perfect case where both are normally distributed
in log space the intrinsic rms, $rms_i$, is related to the measured rms,
$rms_m$, and the average error in line width $\overline{\sigma^2}$ by
\begin{equation} \label{equ_error_corr}
rms_i=\sqrt{rms_{m}^{2} - \overline{\sigma^2}}.
\end{equation}

We are not using the rms in this analysis, instead we take the
68.3\,\% IPV width to
provide an equivalent statistic. Further we do not take
$\overline{\sigma^2}$ to describe the average error in our data since
we find this can be skewed by a small number of objects with very
large errors on their line widths.
To avoid these outliers we take the square of the median error in a
bin to estimate $\overline{\sigma^2}$.

\section{results}

\begin{figure*}
\centering
\centerline{\psfig{file=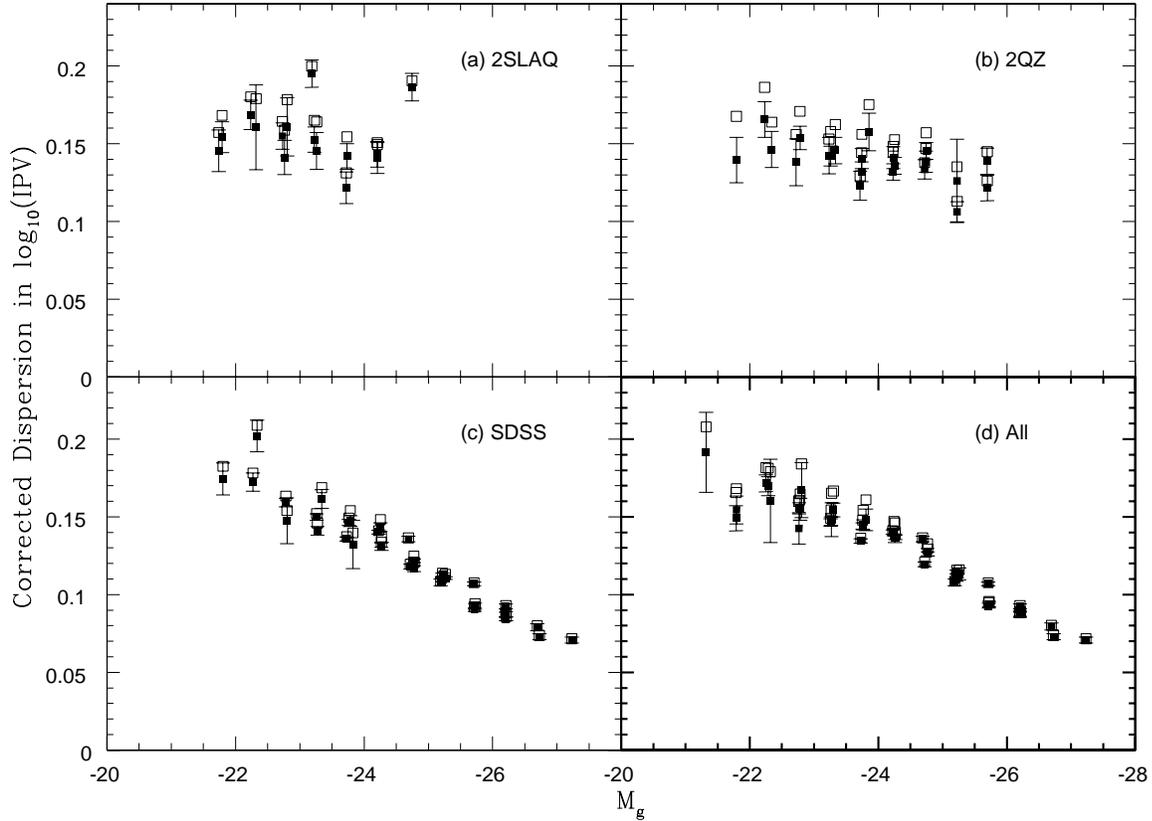,width=16.0cm,angle=-90}}
\caption{Plots of the dispersion in $\log(IPV)$ in
  absolute magnitude bins for the 2SLAQ \empha, 2QZ \emphb\ and SDSS
  \emphc\ samples, as well as the combined sample \emphd. Since we find no 
  redshift dependence in our data we
  plot the results from all redshift bins on the one figure, hence
  more than one point per magnitude interval. The open
  symbols show the dispersion in measured line width before correction
  ($rms_m$; see equation~\ref{equ_error_corr})
  for errors on our measurement, filled symbols with error bars give the
  corrected dispersion ($rms_i$).}
\label{fig_L_disp}
\end{figure*}

We calculate the dispersion in line widths for each $L-z$ bin and
correct by the average error on the line widths in that bin. To ensure
reliable dispersions we only consider $L-z$ bins that contain more
than 40 objects.
Fig.~\ref{fig_L_disp} displays this corrected dispersion vs
luminosity for each of the 2QZ \empha, 2SLAQ \emphb\ and SDSS \emphc\
samples separately, and then for the combined dataset \emphd. Note 
that while we bin our data by redshift we do not see evidence for
redshift dependence in our sample (see section~\ref{sec_z_disp} below)
and so we plot the data from all
redshift bins in Fig.~\ref{fig_L_disp}.

Error bars in the plots
are propagated from the errors on the line width measurements. For
the most part we have a large number of objects in each $L-z$ bin and thus
are dominated not by these random errors but by systematics in the
analysis.

Figs.\ref{fig_L_disp}\empha\ and \emphb\ have considerable scatter and
exhibit little-to-no trend in the plots, although
the shallow trend seen in the 2QZ data is
significant (a Spearman rank test gives $r_s=0.56$ with
$P(r_s)=0.01$).
Fig.\ref{fig_L_disp}\emphc\ shows a strong relation with little scatter
due to the larger numbers in the SDSS sample
(Table~\ref{tab_data}) and its higher quality spectra.

Fig.\ref{fig_L_disp}\emphd\ shows the results when we combine all of our
data. This is heavily dominated by SDSS objects down to absolute magnitudes
$M_g\sim-23$ where the 2QZ and then 2SLAQ samples become important.

Shen et al.~(2007) perform a similar analysis on the SDSS DR5 quasar
sample. They find a consistent level of variance in their data but
report no dependence on luminosity, although inspection of their Fig. 4
does suggest a trend albeit slight. It may be that their following the
McLure \& Dunlop~(2004) prescription for measuring the width of the
\mgii\ line and/or the use of large
luminosity bins in this figure could explain their not finding as
strong a relation as apparent in Fig.\ref{fig_L_disp}.

\subsection{Redshift dependence} \label{sec_z_disp}

\begin{figure}
\centering
\centerline{\psfig{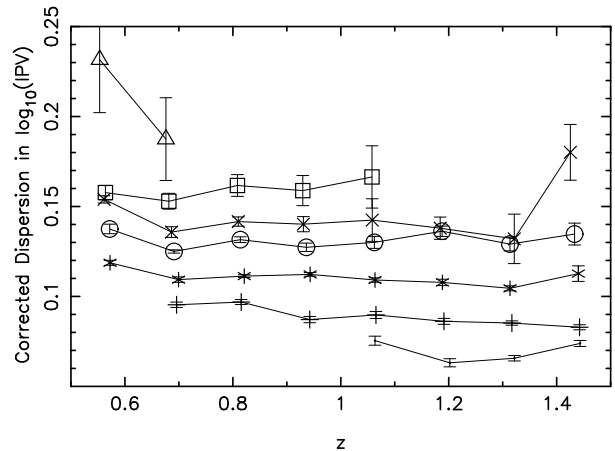}}
\caption{Here we plot the dispersion in $\log(IPV)$ vs. redshift for
  our sample. Luminosity bins are 1\,mag. in width and are centered on
  (top to bottom) $M_g= -21$, $-22$, $-23$, $-24$, $-25$, $-26$ and
  $-27$. Bins with the same luminosity cuts are given the same symbol
  and connected in the plot, and it is evident the dispersion in IPV
  widths in these bins is only very weakly dependent on redshift.}
\label{fig_z_disp}
\end{figure}

We find in Fig.\ref{fig_L_disp} that points from the same luminosity
bin but differing redshift bins lie on top of each other in the plot
suggesting the dispersion in line widths has little-to-no redshift
dependence. This is further illustrated in Fig.\ref{fig_z_disp}. In
this plot we show the dispersion in IPV line width plotted against
redshift for $L-z$ bins. Note that these are not the same $L-z$ bins
as used in the rest of this paper, we have doubled the number of
redshift bins and halved the number of luminosity bins to make the
plot clearer.

In the plot we connect points which are in the same luminosity bin and
it is evident the dispersion in IPV line widths in these bins
has at most a weak redshift dependence. A Spearman rank test
performed only on the three luminosity bins which have full redshift
coverage gives $r_s=-0.14$ significant at only the 50\,\%
level. This is consistent with the changing distribution
of luminosities within each $L$ bin as the luminosity function evolves
with redshift.

\subsection{The effects of the iron template} \label{sec_res_Fe}

As discussed above, the Vestergaard \& Wilkes~(2001) iron template used
in our analysis includes no iron emission directly under the \mgii\
line and we correct for this by adding flux at a level
suggested by theoretical models for QSO iron emission \cite{s+p03}. To
test what effect this has on our results we refit all of our data with
an iron template with no iron emission under \mgii\ and twice as much iron
emission as used in the main analysis. Finally we also try an iron
template where we interpolate between the iron emission peaks at
$\sim2750$ and 2840\,\AA\ (see e.g. Fig. 13 Sigut \& Pradhan 2003);
this template includes
more than five times the flux in the \mgii\ region as in our
primary template.

There is a systematic trend between our results and the amount of iron
emission added under the \mgii\ line. The more emission, the more
scatter we find in our line widths. However, the effect is small. We
find an offset of $\sim0.015$ between the dispersion in log($IPV$)
for objects fit with no iron under the \mgii\ line and twice as much
iron emission as assumed in this analysis, the data presented in
Fig.~\ref{fig_L_disp} lie in between these two. In the case of the
template where we interpolate between the 
iron peaks either side of \mgii\ we find an offset of
$\sim0.08$. While this approach produces a significant
offset this model has no physical
basis and was only implemented to test the extremes of the
distribution.

Furthermore, regardless of which iron template we use in our fitting
the trend in our dispersion analysis is the same. We always find that
more luminous objects show less scatter in line widths than fainter
objects.

\subsection{Completeness and homogeneity}

While the 2SLAQ, 2QZ and primary SDSS QSO surveys all have high
spectroscopic completeness overall ($\sim90$\,\%), towards the faint
flux limit of each survey this drops off. Furthermore, the amalgam of many
separate surveys in the SDSS quasar catalogue and our adding to this
the 2SLAQ and 2QZ spectra as well makes for a very inhomogeneous
sample of objects in terms of selection criteria. However, these
factors are not of great importance in terms of biasing our results or
the trend shown in Fig.\ref{fig_L_disp}.

Incompleteness at faint magnitudes should not have a major affect since
inspection of Fig.\ref{fig_M_lw} shows there is
little variation in the average line width with luminosity. Hence the
dispersion in IPV widths in a luminosity bin is not affected strongly
by the size of a luminosity bin, or the distribution of objects within
a given bin.

Inhomogeneity in selection criteria is also not responsible for the
observed trend in Fig.\ref{fig_L_disp}. This is evident since we find
exactly the same trend in each sample separately. Furthermore we find the same
trend if we only take the primary SDSS quasar sample, albeit with
significantly reduced luminosity coverage.

\section{Discussion} \label{sec_disc}

In Fig.\ref{fig_M_lw} the results of our line fitting are plotted along
with contours of constant SMBH mass and Eddington ratio. While there
is uncertainty in the normalisation of these lines the distribution of
points in the plot relative to them is quite suggestive. The line
width distribution appears to be constrained on at least two sides parallel
to these lines.

Firstly the bottom of the distribution follows the line at
$L/L\ed=1$. Potentially this indicates that the Eddington rate does
represent an upper limit for allowable accretion in QSOs,
constraining our results to lie above this line.

Secondly brighter than $M_g\sim-24.5$ the top of the distribution of line
widths follows the $M\bh=10^{10}$\,\msun\ contour on the plot. This is
likely due to the drop off in the SMBH mass function at high mass. The
space density of SMBHs falls off dramatically for masses above
$10^{9}$\,\msun\ and this creates an upper limit to the distribution
of line widths plotted in Fig.\ref{fig_M_lw}.

Finally, and more speculatively, fainter than $M_g\sim-24.5$ the top
of the line width
distribution appears to be constrained along a line of constant
Eddington ratio around $L/L\ed\sim0.01$. The implication of this
drop off of objects towards low accretion rates is that there
is a preferred level of accretion onto SMBHs for QSOs. This indicates
that the accretion distribution for SMBHs as a whole is bimodal,
with quiescent and active SMBHs occupying distinct areas of mass-accretion
space.

\begin{figure}
\centering
\centerline{\psfig{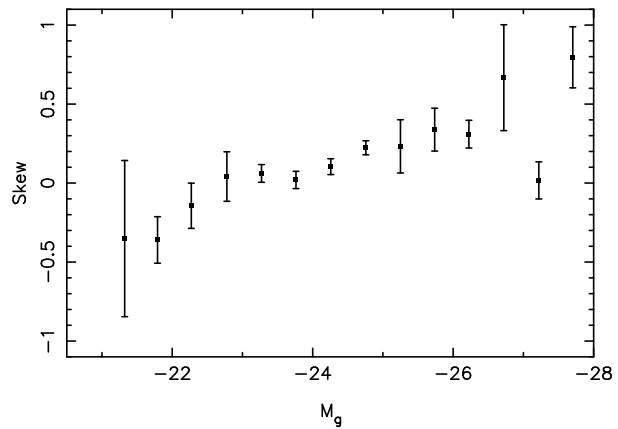}}
\caption{This shows how the skew in the IPV line width distribution is
  dependent on the magnitude of QSOs. The skew is defined by
  equation~\ref{equ_skew} and the error bars are calculated from the
  errors on the IPV width measurements propagated through this equation.}
\label{fig_M_skew}
\end{figure}

As a test as to how the Eddington limit effects the accretion
efficiency distribution we measure the skew of the line width
distribution in magnitude bins. We take the dimensionless skew defined
by the third moment of the IPV width distribution
\begin{equation}
skew = \frac{1}{N} \displaystyle\sum_{i=1}^{n} \left(\frac{IPV_{i}-\overline{IPV}}{\sigma_{IPV}}\right)^{3}.
\label{equ_skew}
\end{equation}
Fig.\ref{fig_M_skew} plots the skew in the IPV width distributions
against the magnitude of the bins.

This figure shows that in more luminous QSOs the line width
distribution consistently has a positive skew. This skew
towards larger line widths and hence lower accretion
efficiencies may be the result of the accretion efficiency
distribution being truncated at the fast end by the Eddington rate.

There is some evidence that this skew reverses for fainter
objects. At these luminosities the mean of the accretion efficiency
distribution is significantly lower than for brighter objects, hence
a cut at the Eddington rate will not have as strong an effect.
The reversal of the trend is likely due to the
underlying mass function of active SMBHs. With many more small SMBHs
there would be a skew towards higher accretion efficiencies and hence
smaller broad line widths at a constant luminosity.

Taken at face value the distribution of points in Fig.\ref{fig_M_lw}
fits well with simple expectations about the QSO SMBH population,
suggesting that the virial mass estimates work relatively well. However,
see section~\ref{sec_rlscat} for a discussion on some concerns with
the virial mass technique which these data highlight. None the less,
this gives us confidence that we can make a meaningful
comparison between our data and theoretical models for activity in the
SMBH population.

\subsection{Comparisons with models}

\begin{figure*}
\centering
\centerline{\psfig{file=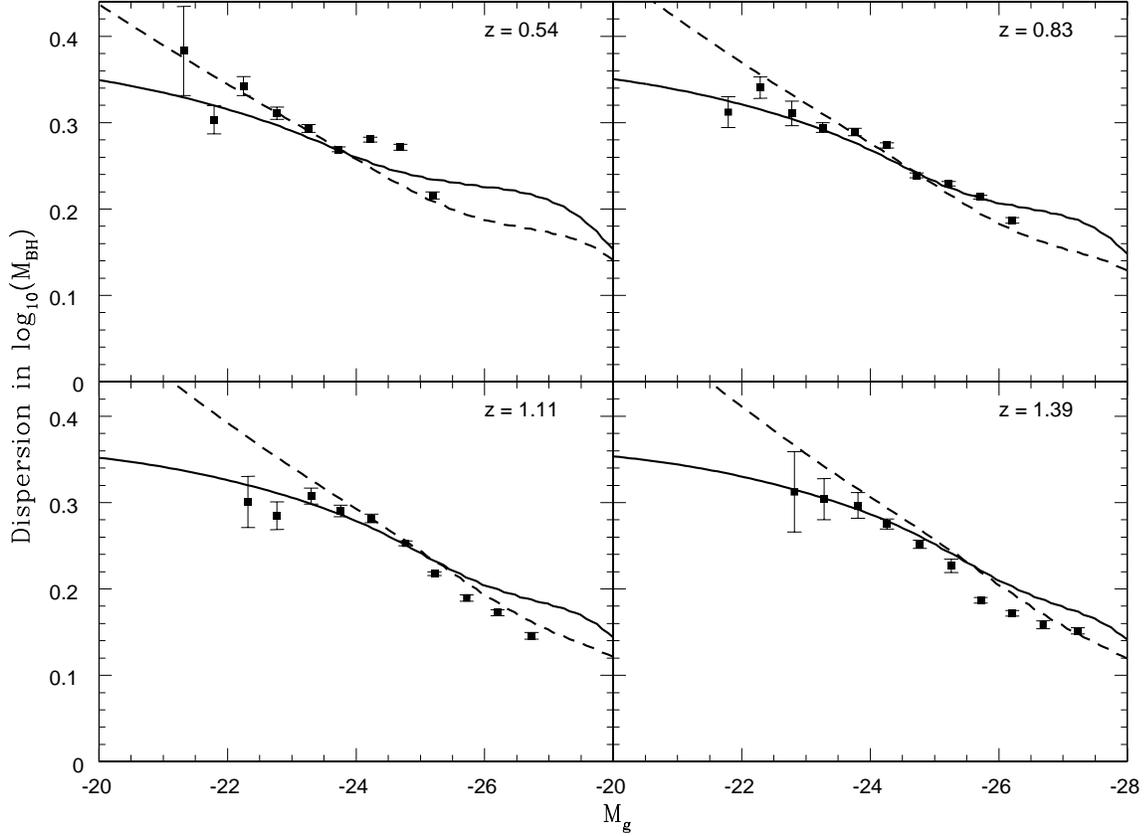,width=16.0cm,angle=-90}}
\caption{Comparison between our data and the models of Hopkins et
al. (2005b). The model lines give the predicted dispersion in
$\log(\mbh )$ as a 
function of luminosity at the mean redshift of the bin (top right of
each panel). The two different curves give the predictions from
differing merger rate functions, either with a sharp cutoff at lower
masses (solid) or a flatter cutoff (dashed). Our measured dispersion in
$\log(IPV)$ is multiplied by 2 for the comparison
(equation~\ref{equ_llw_disp}).}
\label{fig_L_disp_models}
\end{figure*}

In Fig.~\ref{fig_L_disp_models} we compare our results to the model
predictions of Hopkins et al.~(2005b). The models are dependent
on the distributions of galaxy merger rates at a given epoch. This
function is not well constrained by current observations and two
possible realisations are applied. Parametrically the merger rate
function is described by a double power law and the differing models
assume differing slopes for the low mass end. The
first assumes a steep drop off of merging systems towards lower mass
and gives only a narrow
mass range for merging systems at a given redshift. Secondly a broad
range of systems are assumed to be merging at any one time with a flat
slope at the low mass end of the merger function. We label these
models as sharp or flat as describes their low mass slope; the sharp
and flat models are shown as the solid and dashed lines respectively in
Fig.~\ref{fig_L_disp_models}.

We find that our data points match the Hopkins et al.~(2005b) models.
extremely well in general. However, the models exhibit some
evolution with redshift which we do not observe in our data.

\begin{figure}
\centering
\centerline{\psfig{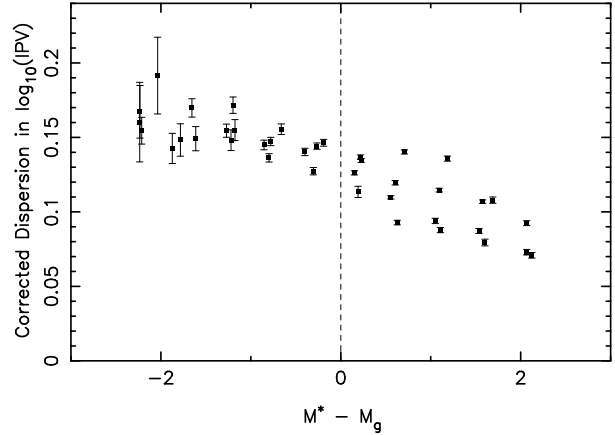}}
\caption{Same data as Fig.\ref{fig_L_disp}\emphd, except the
  dispersion is plotted
  vs. $M^*-M$ rather than $M$. Comparing the scatter evident in this plot
  with Fig.\ref{fig_L_disp}\emphd\ clearly demonstrates the observed
  trend is dependent on the brightness of a quasar, not its position
  on the luminosity function. In this plot we use the quadratic
  parametrisation for $M^*$ given in Croom et al.~(2004).}
\label{fig_Mstar_disp}
\end{figure}

These models are, to an extent, bound to the QSO luminosity
function which shows strong redshift evolution. Over the redshift
range we sample the break in the luminosity function, $M^*$, changes
by more than a magnitude, and the Hopkins
et al. models follow this to an extent. We find no such trend in the
relation shown in Figs.\ref{fig_L_disp}
and~\ref{fig_L_disp_models}. To illustrate this
Fig.\ref{fig_Mstar_disp} shows exactly the same data as
Figs.\ref{fig_L_disp}\emphd\ except the dispersion is plotted against
$M^*-M$, i.e. the object's luminosity relative to the break in the
luminosity function. For this plot we take the quadratic
parametrisation of $M^*$ from Croom et al.~(2004). Comparisons between
Figs.\ref{fig_L_disp}\emphd\
and~\ref{fig_Mstar_disp} clearly show that the observed trend towards
lower dispersion in broad-line width for brighter objects is dependent
on the object's luminosity, not its position on the luminosity function.

The Hopkins et al. models evolve with redshift in a similar sense to
the QSO luminosity function albeit less strongly, hence we observe a
slight discrepancy. However, it was
unlikely that these simple models would be exact in their
determination of the dispersion in $\mbh$. Of more importance is the
trend, which is predicted by the models and well echoed in the data. Our data
show that more luminous QSOs show less scatter in their
\mgii\ line widths than those of lower luminosity, implying they have
less scatter in their black hole masses.

\subsection{Is this trend a selection effect?}

Recently Babi{\'c} et al.~(2007) showed that a broad intrinsic distribution of
Eddington ratios which truncates at the Eddington rate, convolved with
a double power law SMBH mass function for QSOs,
naturally leads to a selection effect such that samples with a
fainter flux limit will find a broader range of Eddington ratios.

To test whether this effect could be responsible for the trend
observed in Fig.\ref{fig_L_disp} we recreate this situation and
compare it with our results. In our analysis we assume that the
distribution of Eddington rates $(L/L\ed =\lambda)$ is log-normal
with a given mean, $\langle\lambda\rangle$, and dispersion, $\sigma_{\lambda}$,
and is truncated at $\lambda=1$. To begin with we assume the SMBH mass
function can
be described as a double power law, and constrain this by convolving
with the Eddington distribution and fitting to the observed luminosity
function (we use the Hopkins, Richards \& Hernquist~2007 $B$-band
luminosity function at redshift 1), producing an (almost) unique
solution for the mass function. We can then derive the dispersion in SMBH
mass at a given luminosity.

\begin{figure*}
\centering
\centerline{\psfig{file=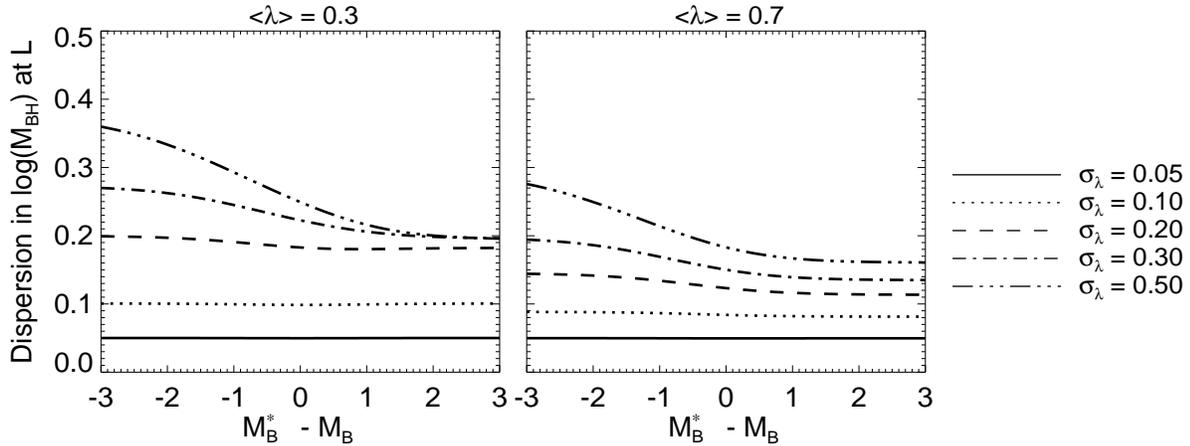,width=16.0cm}}
\caption{These plots illustrate the effect described by Babi{\'c} et
  al.~(2007). The dispersion in SMBH mass at fixed luminosity is
  plotted assuming Eddington ratio distributions with
  means $\langle\lambda\rangle=0.3$ and 0.7, and with a variety of
  widths $(\sigma_{\lambda})$.}
\label{fig_hop_miller}
\end{figure*}

We find that the effect discussed by Babi{\'c} et al.~(2007) is reproduced
in the situation where $\langle\lambda\rangle$ and $\sigma_{\lambda}$ conspire
such that the $\lambda=1$ cut is not too many $\sigma_{\lambda}$'s
away from $\langle\lambda\rangle$; i.e. in situations where the cut has a
significant effect on the Eddington distribution. Hence in situations
where $\langle\lambda\rangle$ is small $\sigma_{\lambda}$ must be 
large to produce the effect. And if $\sigma_{\lambda}$ is
small then $\langle\lambda\rangle$ must be comparable to 1 for a pronounced
effect. Fig~\ref{fig_hop_miller} gives an example of this. We plot the
dispersion in SMBH mass in these models as a function of
luminosity in the case where $\langle\lambda\rangle=0.3$ and 0.7, for
a variety of values of $\sigma_{\lambda}$.

While both of these plots show a decrease in the dispersion in SMBH
mass with luminosity, neither give a good representation of our
results. In the case where $\langle\lambda\rangle=0.3$
$\sigma_{\lambda}$ must be 
so high to get a pronounced gradient that the normalisation of the
relation, in particular at the bright end, is too high when compared
with our results. In the case where $\langle\lambda\rangle=0.7$,
$\sigma_{\lambda}$ can be 
smaller and the normalisation is closer to that observed. However,
the model dispersions flatten a magnitude brighter than
$M^*$, an effect we do not observe.

In all cases the dispersion asymptotes to a constant value above $M^*$
since at these luminosities one is sampling from both a
luminosity function, and by extension SMBH mass function, which follow
a single power law. The Babi{\'c} et al. effect does not occur without
sampling objects from a mass function which is steepening in the log
sense.

This is, however, an artifact of our assigning a double power law for
SMBH mass function. If we assume a Schechter function instead the high
mass end drops off exponentially and the Babi{\'c} et al. effect is
apparent at the higher luminosities sampled. On the other hand with
this model the fit to the observed QSO
luminosity function (which does follow a double powerlaw) is
significantly degraded.


Finally, since these models are tied directly to the luminosity
function (i.e. are relative to $M^*$) they evolve strongly with
redshift. The observed evolution in the QSO luminosity function must be
either due to a similarly evolving active SMBH mass function or an
equivalently varying accretion efficiency distribution. Either 
will strongly affect the Babi{\'c} et al. effect.

As discussed in
section~\ref{sec_z_disp} we find no evidence for a redshift dependence
in our results. And as shown in Fig.\ref{fig_Mstar_disp} the observed
correlation between dispersion in IPV widths and luminosity is
degraded when plotted vs. luminosity relative to $M^*$.

In light of the above discussion it is clear that while the Babi{\'c}
et al.~(2007) effect may well bias our results, it does not accurately
describe both the results presented in Fig.\ref{fig_L_disp} and the
observed luminosity function simultaneously. The
lack of observed evolution in our results may suggest that this effect
is not significant, however, without a better understanding of
the underlying SMBH mass function or accretion efficiency
distribution it is difficult to gauge the magnitude of this effect and
we cannot rule out a significant bias in our data.

\subsection{$\disp(\log(IPV))$ vs $\disp(\log(\mbh))$ and virial
masses revisited} \label{sec_virial_problems}

While we observe a strikingly good agreement between our data and the
the Hopkins et al.~(2005b) we
caution against reading too much into the normalisation of our data
with respect to the dispersion in SMBH mass. In fact a closer
examination of this normalisation raises a question as to just
how accurately the dispersion in $\log(IPV)$ traces that in
$\log(\mbh)$.

In using equation~\ref{equ_llw_disp} we are assuming there are no
other sources of scatter in the QSO SMBH population which will affect
these results. This is almost certainly not the case and we must consider
where other sources of scatter might occur, their magnitude and their
affect on our
results. Shen et al.~(2007) recently performed simulations of the QSO
SMBH population to investigate the biases associated with virial mass
estimations and we follow a similar path to their work.

They describe the QSO and SMBH populations with four variables. The SMBH
mass, the bolometric luminosity (and by extension the
Eddington ratio), the monochromatic luminosity as used in the virial
relation (equation~\ref{equ_virial}) and the FWHM of the spectral
line being analysed (in our case the IPV width of \mgii). We add to
this the radius and velocity dispersion of the
BLR: $r\blr$ and $v\blr$. There are possible sources
of scatter to each of these variables and here we discuss each of
them, their likely magnitude, and their affect on our measurements.

\begin{itemize}
\item {\bf Scatter in luminosity}

The primary source of scatter in luminosity at fixed SMBH mass comes
from the distribution of accretion efficiencies for QSOs. This is
precisely the scatter we are trying to measure in this work and is not
a contaminant to our calculations.

The only other sources of scatter to the luminosity are photometric
errors, host galaxy light and extinction. Each of these would add
extra scatter into the luminosities we measure independently from any
virial mass equations. Hence this scatter would not propagate
through to the line widths we measure in the way scatter in accretion rate
would. This has the unique effect of making the SMBH mass
distribution we measure at a given luminosity narrower relative to the
intrinsic mass distribution (see Shen et al.~2007).

However, these sources of scatter can only play a minor role in our
measurements and calculations. Photometric errors are $\sim0.1$\,mag.
in the $b_{\rm J}$-band used by the 2QZ ($\sim0.04$\,dex in $L$;
$\sim0.02$\,dex in $\mbh$) and
smaller for the SDSS observations. Host galaxy emission and extinction
will only affect the fainter objects in our sample, and even at these
luminosities will not be a major factor. Hence these sources of
scatter will not be significant to our calculations.

Shen et al. also include a scatter due to variations in continuum
shape when converting bolometric luminosities to monochromatic
luminosities. They quote a value for this of $\sim0.1$\,dex
translating to $\sim0.05$\,dex scatter in $\mbh$ and so again this is
not of great importance to our analysis.

\item {\bf Scatter in $r\blr$}

The \rl\ relation shows that the radius of the BLR around QSOs is
driven by the luminosity of the source. Hence any scatter in
the intrinsic luminosity of the source (i.e. not due to photometric
errors/host galaxy emission/extinction) will also scatter
$r\blr$. Furthermore the \rl\ relation has
been observed to exhibit intrinsic scatter of its own at a level of
40\,\% \cite{kasp05} translating to at least 0.15\,dex, and this
intrinsic scatter will further broaden the $r\blr$ distribution. This
extra scatter will propagate through to the IPV width distribution we
measure, and
will bias our final measurement for the scatter in $\mbh$ high by the same
factor. 0.15\,dex of scatter in $\mbh$ is significant to our
discussion and we expand on the implications of this in
section~\ref{sec_rlscat}.

\item {\bf Scatter in $v\blr$}

The distribution of velocity dispersions will have all of the above
scatters folded into it apart from the scatter in luminosity due to
photometric errors/host galaxy emission/extinction. If we take $v\blr$
to be the virial velocity of the BLR then there will be
no additional sources of scatter which affect this quantity since the
virial equation (eqn.~\ref{equ_virial_real}) is exact.

However, other (non-virial) factors may effect the velocity of the BLR.
Shen et al. include an additional scatter to the line width distribution
due to non-virial velocities in the BLR. This will
bias the scatter we measure in IPV widths (and hence $\mbh$)
high, although in this work we do not attempt to investigate this further.

\item {\bf Scatter in IPV widths}

Beyond all of the scatters which affect $v\blr$, the IPV width
distribution we measure can be effected by further sources of
scatter. Firstly errors on our measurements will artificially broaden
the measured IPV width distribution, and we account for this in our
dispersion analysis (see section~\ref{sec_disp_anal} and
equation~\ref{equ_error_corr}).

Secondly the line width we measure may not accurately describe the
velocity of the BLR. Selective absorption (e.g. the model
proposed for the \civ\ BLR by Richards et al. (2002) and/or orientation
dependent line widths would further broaden the measured IPV width
distribution. There is little evidence for absorption in the \mgii\
line for non-BAL QSOs and do not attempt to model its effects here.
The potential effects of orientation on our measurements are discussed
in section~\ref{sec_blr_geom} below.

\end{itemize}

Of these sources of scatter, only the uncorrelated scatter
in luminosity can bias the measured dispersion in IPV widths
low. All of the other sources of scatter propagate
through to the IPV distribution and will bias the results high.

%

Little is known about the magnitude of many of the sources of scatter
discussed above,
but most can be ruled out as negligible when compared to the observed
dispersion in IPV widths. The only source of scatter which is known to be
comparable to that measured for the IPV widths is the intrinsic scatter
in the \rl\ relation.

\subsubsection{Implications of intrinsic scatter in the \rl\ relation}
\label{sec_rlscat}

Equation~\ref{equ_llw_disp} assumes that there is no variation in
$r\blr$ within a luminosity bin. This is not the case, and the
distribution of line widths in a given luminosity-redshift bin
is not directly analogous to the
intrinsic distribution of SMBH masses. Instead the distribution of
line widths is a convolution of the black hole mass distribution
and the distribution of BLR radii in a given bin (as well as our error
distribution; equation~\ref{equ_error_corr}).

With this in mind a better estimate of the true dispersion in $\mbh$
is the calculated dispersion in $\log(IPV)$ ($\times2$;
equation~\ref{equ_llw_disp}) with both the
dispersion in luminosities in a single bin ($\times\alpha$;
equation~\ref{equ_virial}) and the intrinsic scatter in
the \rl\ relation subtracted from it in quadrature.

The scatter in
$\log(L)$ for 0.5\,mag. bins will be $\sim0.1$\,dex in the faint bins,
and decrease to brighter luminosities as the luminosity function
steepens. This translates to $<0.05$\,dex of scatter in $\mbh$ and
hence is relatively unimportant to this discussion.

On the other hand the intrinsic scatter in the \rl\ relation is not
negligible with
respect to our results. Since the radius of the
BLR is dependent on the luminosity of the quasar (and not the other
way around), any intrinsic scatter in this relation will also tend to
bias our results high. Due to the relatively
small number of objects with good reverberation data the scatter in
the \rl\ relation is not
definitively constrained, and not defined at all for the
brighter luminosities in our sample. However, Kaspi et
al. (2005) do measure this for the \hbeta\ BLR with 35 of the best
mapped AGN and quote a value of $\sim40$\,\%
for the intrinsic scatter in $r\blr$. This corresponds
to at least $\sim0.15$\,dex in the log (0.17 if the error is
propagated as in equation~\ref{equ_error_prop}), and is greater than
the scatter we find in SMBH mass in the brightest magnitude bins.

This leaves a dilemma: Either there is significantly less
intrinsic scatter in the \rl\ relation than that quoted by Kaspi et
al. (at least at high $L$), or there is an intrinsic problem with our
approach to finding the scatter in $\mbh$.

The 40\,\% scatter quoted by Kaspi et al.~(2005) is not directly
applicable to our analysis for two reasons: 1) The \rl\ relation
studied in that work is for the \hbeta\ emission region not \mgii\ and 2)
The \rl\ relation is only defined for relatively faint Seyfert~1-type
objects, while the lowest scatter we find is for the most luminous quasars
in our sample. In effect our use of the Kaspi et al. value for the
scatter in this relation equates to an extrapolation over $\sim2$
orders of magnitude in luminosity.


In their analysis Kaspi et al. suggest possible sources of scatter in
the \rl\ relation such as intrinsic reddening, contributions by the
host galaxy, or effects of variability due to non contemporaneous
observations. Of these, galaxy contamination and potentially reddening
will play a smaller role at higher luminosities. Bentz et al.~(2006)
used $HST$ imaging to correct a subset of the objects in Kaspi et al.'s
sample for host galaxy emission and found that the intrinsic scatter
in the \rl\ relation could be as low as $\sim30$\,\%. At higher
luminosities where host contamination would be lessened it is possible
that the scatter is even less.


If there were no scatter in the \rl\ relation our data imply there is
only $\sim0.14$\,dex scatter in $\mbh$ in luminous QSOs. This
may be plausible considering that at the extreme luminosities we are
considering here we would expect most quasars to be radiating at or
around their Eddington luminosity.

An alternative interpretation may be that the
dispersion in $\log(IPV)$ does not properly represent the dispersion
in $\log(\mbh)$. That is, QSO broad lines do not show enough variation to
account for the expected variation in $\mbh$ when accounting for the
intrinsic scatter in the \rl\ relation.

This raises a question as to the validity of the analysis
performed here, and the virial method for estimating $\mbh$.

There is evidence for a virialised BLR. Time lags between
continuum and emission line variations have been observed 
to be faster in the wings as opposed to the core of lines
\cite{koll96,o+p02,koll03}. And the same authors also find that in
objects which have had more than one line mapped their time delays appear to
follow a virial $(\tau\propto V\blr^{-2})$ relation.
In addition, studies comparing virial masses to bulge velocity
dispersion have shown a correlation analogous to that found in local
quiescent galaxies \cite{onk04,woo06} indicating that the virial
method works as a tracer of $\mbh$.

An explanation could be that the BLR of QSOs is not
{\emph{wholly}} virialised. And there is a significant component of
the BLR which shows very little object-to-object velocity variation.
If this were the case virial motion would still be the
primary cause of the variation between broad line widths in QSOs and
observations such as the \ms\ relation would be reproduced. However, globally
we would not see a range of line widths comparable to the range of
black hole masses.

Alternatively, and more controversially, it may be that the BLR is not
virialised at all and the
virial relations only prove accurate due to their luminosity
dependence. The SMBH mass of a QSO and its luminosity must be strongly
covariant and the ability
of virial estimators to determine $\mbh$ may be due to this simple
relation. The line width term may be redundant. Indeed we (and
other authors e.g. Corbett et al.~2003; Shen et al~2007) find the
average line width changes very little over the luminosity range of our
sample. To be consistent with the virial mass estimators this requires
that both SMBH mass and Eddington ratio for QSOs vary as luminosity to
the power of $\sim1/2$. A conspiracy which must be viewed with some caution.

The reason for the lack of observed scatter in broad line widths
is uncertain, but it is apparent that this observation has
consequences for black hole mass estimation in AGN. While this may imply
that virial masses are not necessarily unbiased
estimators of $\mbh$ there is evidence from other studies that they
can be used as an indicator of the mass of a black hole.

\subsection{Constraining BLR geometry} \label{sec_blr_geom}

The low dispersion in broad-line width found for luminous QSOs is also a
strong constraint on the velocity field of the BLR.

In all but the case of a spherically symmetric BLR
the velocity dispersion we measure from a spectrum will depend
on the viewing angle to the QSO. Given a model for the BLR we can
calculate the expected dispersion in measured line widths due to
variations in observing angle. If
the expected dispersion is greater than we find in our data we can
rule out that model for the BLR. Since we take no account
of the myriad of other sources of scatter in the line width
distribution, this equates to a very strong constraint on a given
BLR model.

\begin{figure}
\centering
\centerline{\psfig{file=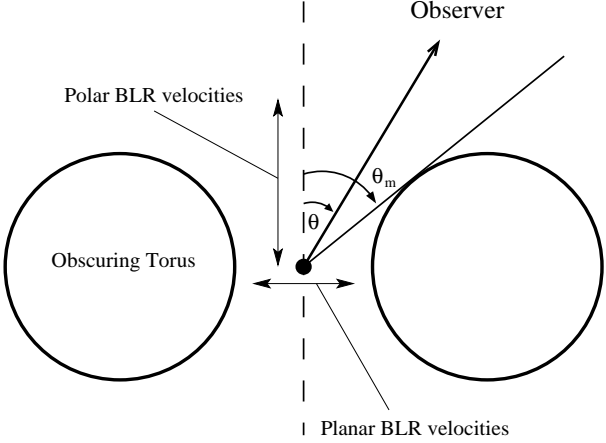,width=8.0cm}}
\caption{A sketch showing the assumed geometry of an AGN. The central
  SMBH and BLR are surrounded by an obscuring torus which constrains
  the allowed observing angle ($\theta$) to be less than some opening
  angle ($\theta_m$)}
\label{fig_AGN_schem}
\end{figure}

\subsubsection{Pure planar/polar BLRs}

The first models we consider are planar velocity fields
in which all the velocities are confined to the $z=0$ plane
(cylindrical coordinates), and polar fields in which all the
velocities are in the $\hat{z}$ direction. In these cases we expect our
measured IPV widths to be modified by

\begin{tabular}{p{1cm}r}
Planar: & $\log(IPV) \propto \log(\sin(\theta))$ \\
Polar:  & $\log(IPV) \propto \log(\cos(\theta))$ \\
\end{tabular}

\noindent
where $\theta$ is our observing angle to the QSO (see
Fig.\ref{fig_AGN_schem}). Assuming an opening angle ($\theta_m$),
which we take to be constrained by an obscuring torus
(Fig.\ref{fig_AGN_schem}), we can calculate what the expected dispersion
in measured line widths would be due solely to orientation effects for
planar and polar velocity fields. 

\begin{figure*}
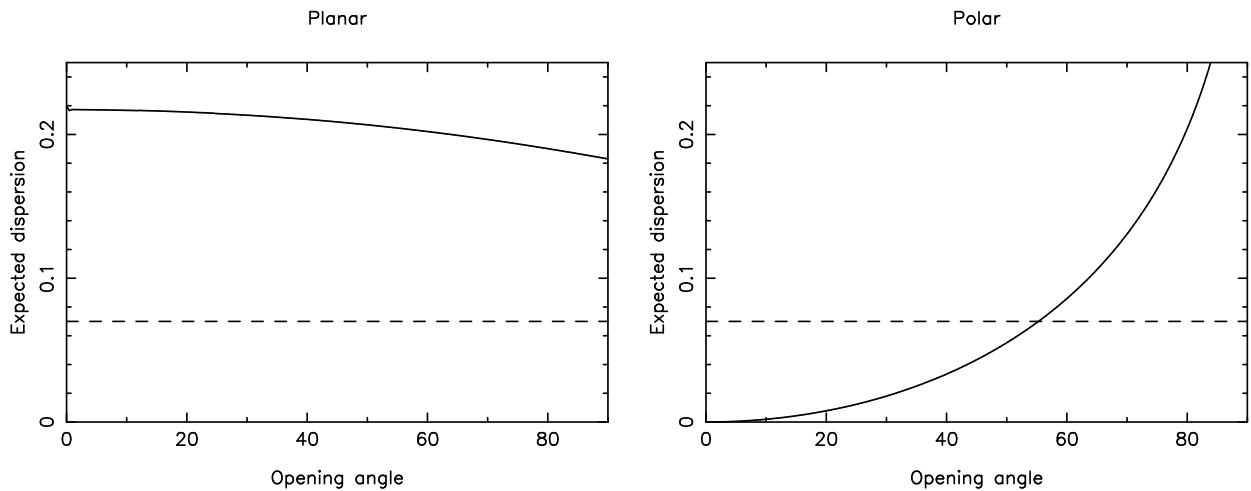

\centering
\centerline{\psfig{file=oang_100_planar.ps,width=8.0cm,angle=-90}\hspace{0.5cm}\psfig{file=oang_100_polar.ps,width=8.0cm,angle=-90}}
\caption{These plots show how the dispersion in measured line
  width is dependent on
  the opening angle of the source in the special cases where either
  all of the BLR velocities are in the planar or
  polar directions. On each plot the dashed line shows the ${\rm
  dispersion}=0.07$\,dex level, the lowest dispersion in line width we
  measure in our data.}
\label{fig_plane_pole}
\end{figure*}

Fig.\ref{fig_plane_pole} shows how much scatter in line widths we expect
in each case. On both
plots we indicate with a dashed line 0.07\,dex dispersion which is the
lowest we measure in our sample (Fig.\ref{fig_L_disp}).

In the planar case the expected dispersion is $\sim0.2$\,dex for all opening
angles. We find somewhat less dispersion in IPV widths than this in
all but two of our $L-z$ bins. Hence the velocity field
the BLR cannot solely be planar but must include some other component.

For the polar velocity field the expected dispersion is a strong
increasing function of $\theta_m$. For modest opening angles the
resulting dispersion is low and it is not until
$\theta_m\sim55\,^{\circ}$ that this becomes comparable to the
dispersions we measure in our data. Hence if the BLR is characterised
by polar flows, the opening angle to luminous quasars must be less than
this value.

\subsubsection{Planar/Polar BLRs with a random/spherically symmetric component}

It is not likely that in any model for the BLR all the
velocities are confined to a single plane/direction. Instead there will
always be some random component to the velocity field and we
include this in our models with the following parametrisation:

\begin{tabular}{p{1cm}l}
Planar: & $\log(IPV) \propto \log(a\sin(\theta) + (1-a))$\\
Polar:  & $\log(IPV) \propto \log(a\cos(\theta) + (1-a))$\\
\end{tabular}

\begin{figure*}
\centering
\centerline{\psfig{file=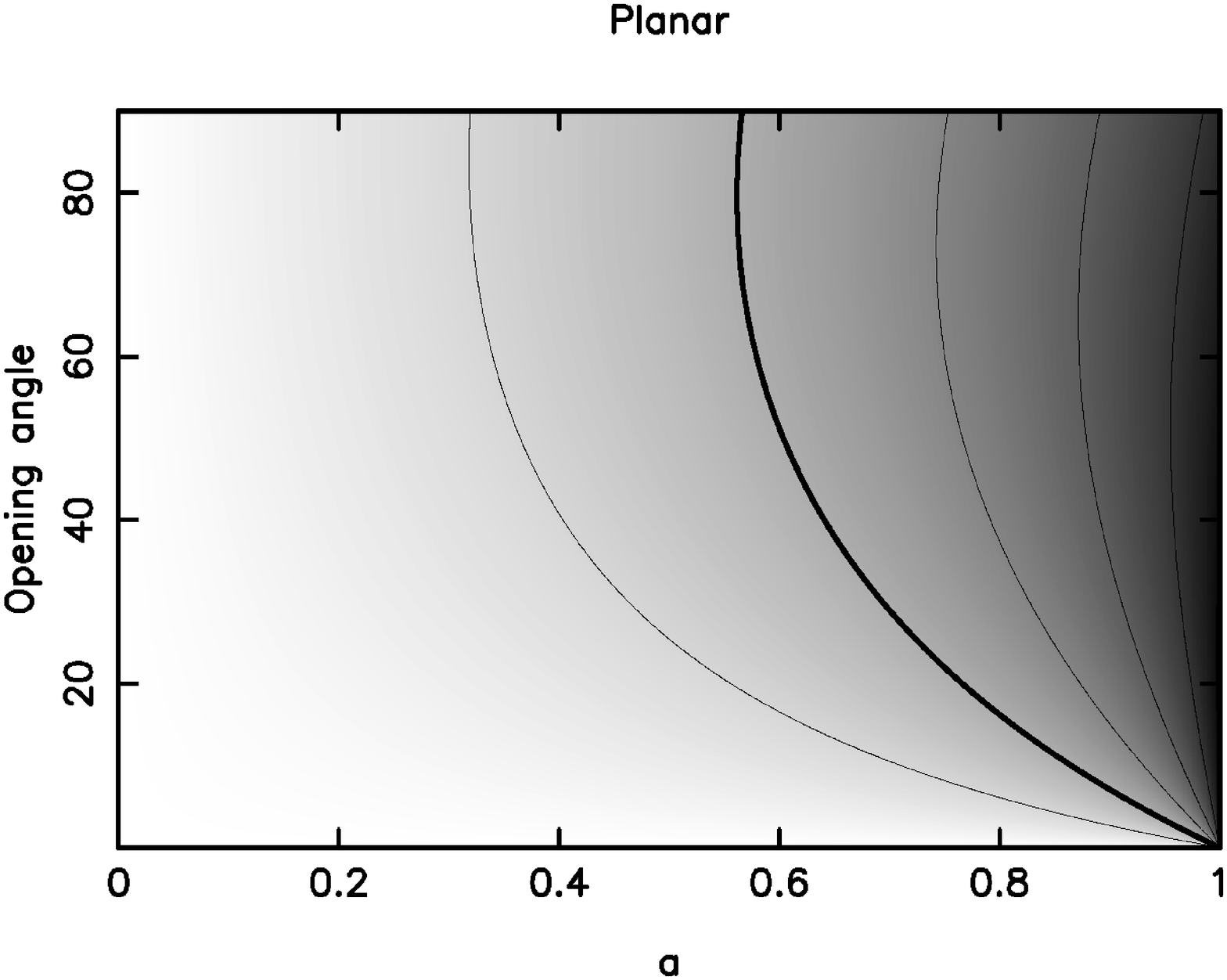,width=8.0cm,angle=-90}\hspace{0.5cm}\psfig{file=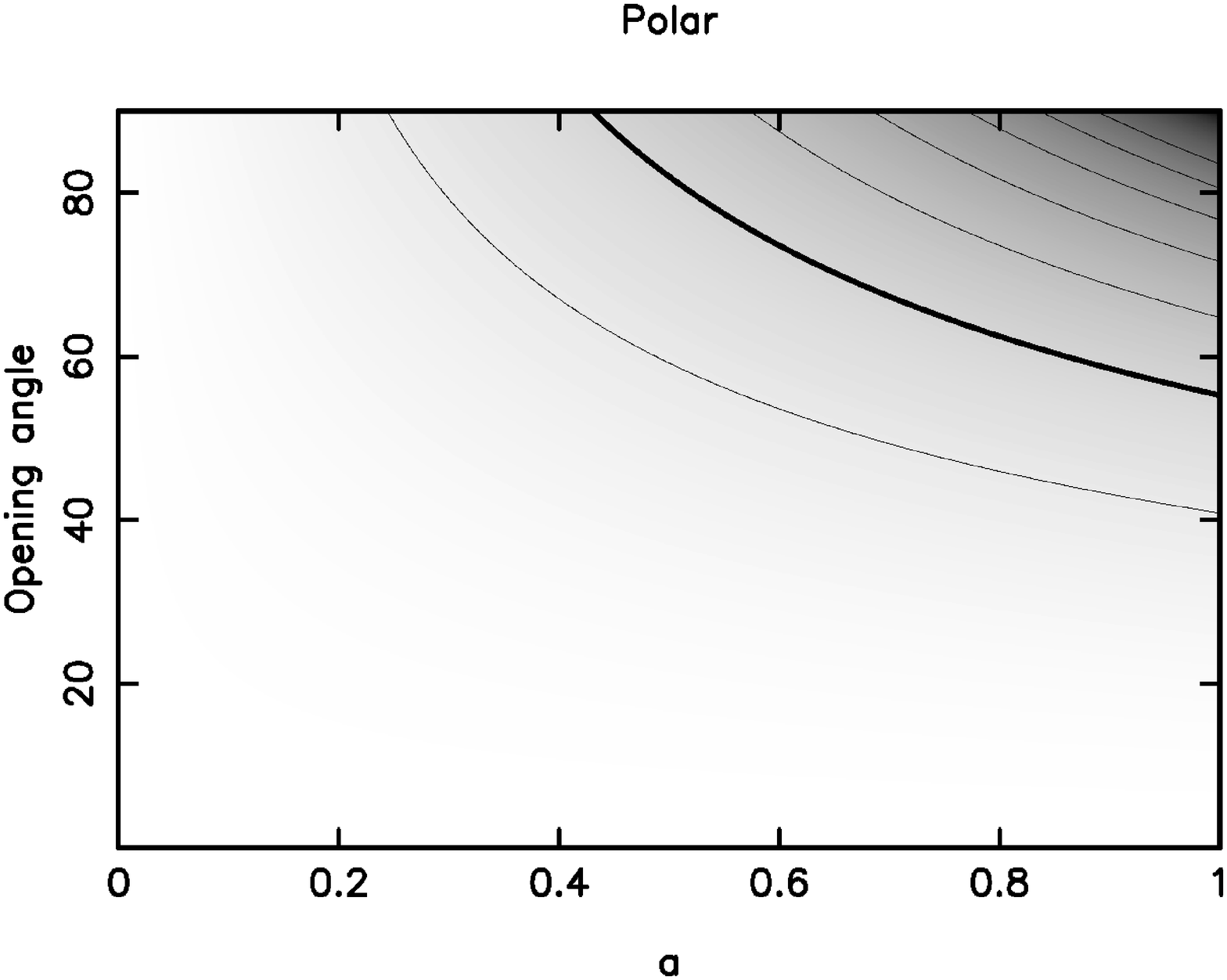,width=8.0cm,angle=-90}}
\caption{Here we demonstrate how adding a spherically symmetric
  component to the 
  BLR velocity field affects the expected dispersion in broad line
  widths. Again we assume that all of the other velocities are either in
  the planar or polar directions. In our parameterisations $a\sim0$
  indicates a BLR dominated by a spherically symmetric velocity field,
  while $a\sim1$ indicates no symmetric component equivalent to the models
  shown in Fig.\ref{fig_plane_pole}. The grey scale and contours
  indicate the expected dispersion. Contour levels are in increments
  of 0.035\,dex and the second contour (heavy) shows the 0.07\,dex
  scatter we measure for the most luminous objects in our sample. The
  parameter space above this is inconsistent with our analysis.}
\label{fig_disk_wind}
\end{figure*}

In these models the trigonometric term represents a geometrically constrained
component to the BLR velocity, and the $(1-a)$ term represents a
random/spherically symmetric
component. Hence a model with $a=1$ represents the pure planar/polar
cases outlined above, and if $a=0$ the BLR velocity field is
spherically symmetric. In these cases, given values for $a$ and
$\theta_m$ we can again calculate the expected dispersion in
$\log(IPV)$ to compare with our results. These calculations are
illustrated in Fig.\ref{fig_disk_wind}.

In this figure the grey scale indicates the expected scatter at a
point in the $a-\theta_m$ plane. The contours are at increments of
0.035\,dex, and so the second contour (heavy line) represents the
0.07\,dex scatter we observe for the most luminous QSOs in our
sample. Hence, for these luminous objects the region of parameter space
above the solid line is ruled out in our analysis.

In the planar case, for which the constraints are stronger,
taking a believable value for the opening angle
($\sim45^{\circ}$) our data show that roughly half of the
contribution to the velocity field of the BLR must come from a symmetric
component.

\subsubsection{A hybrid BLR}

As a final model we consider a BLR with both planar and polar
components to the velocity field as well as a spherically symmetric
component. We model this with:

$\log(IPV) \propto \log(a\sin(\theta) + b\cos(\theta) + (1-a-b))$

\begin{figure*}
\centering
\centerline{\psfig{file=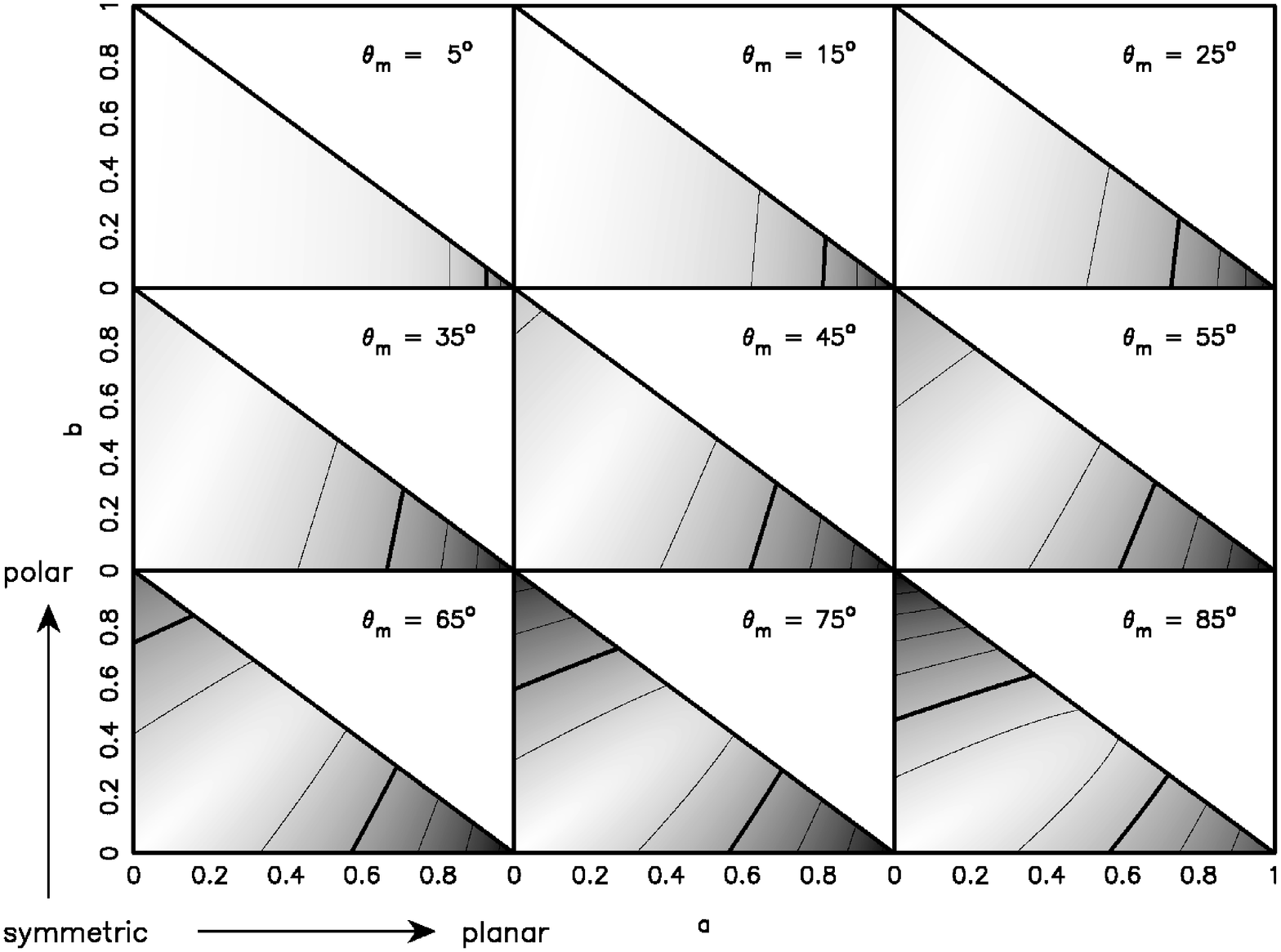,width=16.0cm,angle=-90.}}
\caption{These plots show the expected dispersion in measured line
  widths for our composite BLR model which include planar, polar
  and symmetric components. Each plot shows the expected dispersion
  (grey scale + contours as in Fig.\ref{fig_disk_wind}) for a given AGN
  opening angle ($\theta_m$). In our parametrisation as $a$ increases
  the BLR is more planar, and as $b$ increases more polar. (0,0)
  indicates a totally symmetric velocity field while all models on the
  line $a+b=1$ have no symmetric component.}
\label{fig_disk+wind}
\end{figure*}

\noindent
And so for a given $(a,b,\theta_m)$ we can calculate the expected
dispersion in measured line width. Fig.\ref{fig_disk+wind} shows the
expected dispersion in
the $a-b$ plane for differing opening angles. In each plot the origin
$(a=b=0)$ represents a spherically symmetric BLR and the line $a+b=1$
represents a BLR with no symmetric component. As one moves from left to
right in each plot the BLR becomes more planar, and from bottom to
top the velocity field is more polar.

As expected, for small opening angles our data are not strong
constraints on the velocity field of the BLR, and only the most
disk-like models are rejected. For larger opening angles there is a
larger region of the parameter space ruled out by our data,
predominantly the very planar or very polar models.

\section{Conclusions}

We have measured the 50\,\% IPV width of the \mgii\ line in QSO spectra
from the SDSS, 2QZ and 2SLAQ surveys and find a strong correlation between
the dispersion in IPV widths and the optical luminosity of QSOs. If we
assume there exists a virial relation of the form of
equation~\ref{equ_virial} this implies that there is an equivalent
reduction in the dispersion in $\mbh$. On face value this is in excellent
agreement with models for the QSO population proposed by Hopkins et
al.~(2005b).


However, the remarkably low scatter we find in
our IPV width measurements, in particular for the more luminous objects, has
implications as to the validity of virial mass estimators. We find
less scatter in IPV
widths of luminous QSOs than is intrinsic to the \rl\ relation. While
the \rl\ relation is not defined for these very luminous objects
these results are at odds with the practise of virial black
hole estimation. Possible explanations for this observation include
very low scatter in the \rl\ relation for bright QSOs or a BLR which
is not fully virialised. In either case it is clear we are yet to gain
a full understanding of the process of virial black hole mass
estimation and one must be cautious when performing or interpreting
these estimates.

Finally we show that the observed scatter in IPV widths can be used to
constrain models for the velocity field of the BLR. We show how
variations in observing angle to sources affects the scatter in
measured IPV widths for a series of simplified geometries and derive
constraints on these models from our data. Perhaps of
most interest is the rejection of a pure planar BLR regardless of the
assumed opening angle to the source.

A table containing the \mgii\ IPV widths for the quasars studied in
this work is available through the 2SLAQ website ({\sc www.2slaq.info}).

\section{Acknowledgements}

We would like to thank all our colleagues who gave useful input into
this work. In particular we thank Yue Shen for some enlightening
discussions. Furthermore SMC acknowledges the support of an Australian
Research Council QEII Fellowship and an J G Russell Award from the
Australian Academy of Science.

In addition the authors would like to  thank the SDSS project from
which much of the data in this paper was obtained.
Funding for the SDSS and SDSS-II has been provided by the Alfred
P. Sloan Foundation, the Participating Institutions, the National
Science Foundation, the U.S. Department of Energy, the National
Aeronautics and Space Administration, the Japanese Monbukagakusho, the
Max Planck Society, and the Higher Education Funding Council for
England. The SDSS Web Site is http://www.sdss.org/. 

The SDSS is managed by the Astrophysical Research Consortium for the
Participating Institutions. The Participating Institutions are the
American Museum of Natural History, Astrophysical Institute Potsdam,
University of Basel, University of Cambridge, Case Western Reserve
University, University of Chicago, Drexel University, Fermilab, the
Institute for Advanced Study, the Japan Participation Group, Johns
Hopkins University, the Joint Institute for Nuclear Astrophysics, the
Kavli Institute for Particle Astrophysics and Cosmology, the Korean
Scientist Group, the Chinese Academy of Sciences (LAMOST), Los Alamos
National Laboratory, the Max-Planck-Institute for Astronomy (MPIA),
the Max-Planck-Institute for Astrophysics (MPA), New Mexico State
University, Ohio State University, University of Pittsburgh,
University of Portsmouth, Princeton University, the United States
Naval Observatory, and the University of Washington.  

The authors would also like to thank all the present and former staff of the
Anglo-Australian Observatory for their work in building and operating
the 2dF facility.  The 2QZ and 2SLAQ are based on
observations made with the Anglo-Australian Telescope and the UK
Schmidt Telescope as well as the Sloan telescope.

We would also like to thank all of the good people
at the University of Sydney for their help, their advice and their
on-going support. And the staff at the Harvard-Smithsonian Centre for
Astrophysics for their significant contributions to this work.

\end{document}